\documentclass[twocolumn,trackchanges]{aastex63}

\received{July 6, 2020}
\revised{July 24, 2020}
\accepted{July 24, 2020}
\submitjournal{AAS Journals}

\shorttitle{The Most Metal-poor Stars in the Inner Bulge}
\shortauthors{Reggiani et al.}

\begin{document}

\title{The Most Metal-poor Stars in the Inner Bulge\footnote{This paper
includes data gathered with the 6.5-meter Magellan Telescopes located
at Las Campanas Observatory, Chile.}}

\correspondingauthor{Henrique Reggiani}
\email{hreggiani@jhu.edu}

\author[0000-0001-6533-6179]{Henrique Reggiani}
\affiliation{Department of Physics and Astronomy, Johns Hopkins
University, 3400 N Charles St., Baltimore, MD 21218, USA}

\author[0000-0001-5761-6779]{Kevin C. Schlaufman}
\affiliation{Department of Physics and Astronomy, Johns Hopkins
University, 3400 N Charles St., Baltimore, MD 21218, USA}

\author[0000-0003-0174-0564]{Andrew R. Casey}
\affiliation{School of Physics \& Astronomy, Monash University, Wellington
Road, Clayton 3800, Victoria, Australia}
\affiliation{ARC Centre of Excellence for All Sky Astrophysics in 3
Dimensions(ASTRO 3D), Canberra, ACT 2611, Australia}

\author[0000-0002-4863-8842]{Alexander P. Ji}
\altaffiliation{Hubble Fellow}
\affiliation{The Observatories of the Carnegie Institution for Science,
813 Santa Barbara St., Pasadena, CA 91101, USA}

\begin{abstract}

\noindent
The bulge is the oldest component of the Milky Way.  Since numerous
simulations of Milky Way formation have predicted that the oldest
stars at a given metallicity are found on tightly bound orbits,
the Galaxy's oldest stars are likely metal-poor stars in the inner
bulge with small apocenters (i.e., $R_{\mathrm{apo}}\lesssim4$ kpc).
In the past, stars with these properties have been impossible to find
due to extreme reddening and extinction along the line of sight to the
inner bulge.  We have used the mid-infrared metal-poor star selection of
\citet{schlaufman2014} on Spitzer/GLIMPSE data to overcome these problems
and target candidate inner bulge metal-poor giants for moderate-resolution
spectroscopy with AAT/AAOmega.  We used those data to select three
confirmed metal-poor giants ($[\mathrm{Fe/H}]=-3.15,-2.56,-2.03$) for
follow-up high-resolution Magellan/MIKE spectroscopy.  A comprehensive
orbit analysis using Gaia DR2 astrometry and our measured radial
velocities confirms that these stars are tightly bound inner bulge stars.
We determine the elemental abundances of each star and find high titanium
and iron-peak abundances relative to iron in our most metal-poor star.
We propose that the distinct abundance signature we detect is a product
of nucleosynthesis in the Chandrasekhar-mass thermonuclear supernova of
a CO white dwarf accreting from a helium star with a delay time of about
10 Myr.  Even though chemical evolution is expected to occur quickly in
the bulge, the intense star formation in the core of the nascent Milky
Way was apparently able to produce at least one Chandrasekhar-mass
thermonuclear supernova progenitor before chemical evolution advanced
beyond $[\mathrm{Fe/H}]\sim-3$.

\end{abstract}


\section{Introduction}\label{intro}

Since galaxies form from the inside-out, the bulge is the oldest major
component of the Milky Way. While
there are at least six physical processes that may have contributed
to the growth of the Milky Way's bulge \citep[e.g.,][]{barbuy2018},
it is statistically implausible that the earliest stage of Milky Way
formation failed to contribute to the bulge's stellar population
at some level.  Indeed, numerical simulations of Milky Way-analog
formation have consistently shown that the metal-poor stars in the
inner few kpc of a Milky Way-like galaxy are often the oldest stars
in the dark matter halo hosting the galaxy.\footnote{See for example
\citet{diemand2005}, \citet{scannapieco2006}, \citet{brook2007},
\citet{salvadori2010}, \citet{gao2010}, \citet{tumlinson2010},
\citet{ishiyama2016}, \citet{starkenburg2017b}, \citet{griffen2018},
and \citet{sharma2018}} At the same time, the early chemical evolution
of the bulge is expected to differ significantly from that of the halo
and surviving dwarf galaxies \citep[e.g.,][]{kobayashi2006}.  For these
reasons, the exploration of this first stage of Milky Way formation using
the chemical abundances of ancient metal-poor stars in the inner Galaxy
has long been a goal of Galactic archaeology.

Some of the expected differences in the chemical
evolution of the bulge and halo have already been observed
\citep[e.g.,][]{kobayashi2006,cunha2006,johnson2012,bensby2017,gargiulo2017,barbuy2018}.
The star formation rates in the event (or events) that lead to the
formation of Milky Way's ``classical bulge'' component are thought
to have been very high.  This intense star formation in a metal-poor
environment would have produced many otherwise uncommon stellar systems,
perhaps including the progenitors of relatively rare classes of supernovae
(e.g., hypernovae, spinstars, thermonuclear supernovae on core-collapse
supernovae timescales, etc.).  The existence of metal-poor stars in the
inner bulge despite this intense star formation requires the accretion
of unenriched gas on short timescales as expected in the denser and more
gas-rich $z \gtrsim 2$ Universe.  Both of these differences between the
bulge and the halo/surviving dwarf galaxies---frequent contributions from
rare supernovae and the signature of ongoing accretion of unenriched
gas---should be apparent in comparisons of the detailed elemental
abundances of the most metal-poor stars in the inner bulge, halo, and
surviving dwarf galaxies.

While metal-poor stars in the inner Galaxy were historically
difficult to separate from the much more numerous metal-rich
stars in the bulge, significant progress has been made in the
last few years.  Several groups have discovered metal-poor
giants in the outer bulge photometrically using ultraviolet
or mid-infrared photometry or spectroscopically in multiplexed surveys
\citep[e.g.,][]{garcia-perez2013,ness2013,casey2015,howes2015,howes2016,lamb2017,lucey2019,arentsen2020}.
In spite of this recent progress in the outer bulge, it has been
impossible to study in detail or even find metal-poor stars with
$[\mathrm{Fe/H}] \lesssim -2.0$ in the inner bulge (i.e., $|l,b| \lesssim
4^{\circ}$) due to the extreme extinction and reddening in that direction.

In this paper, we have used the infrared-only metal-poor star selection of
\citet{schlaufman2014} to discover the most metal-poor stars known in the
inner bulge.  Follow-up high-resolution optical spectroscopy has revealed
that at $[\mathrm{Fe/H}] \sim -3$, the inner bulge has high silicon and
iron-peak abundances relative to iron when compared to giant stars in
the halo or surviving dwarf galaxies.  We attribute these differences
to the occurrence of at least one Chandrasekhar-mass thermonuclear
supernova that occurred early in the bulge's chemical evolution
on a timescale comparable to core-collapse supernovae.  We outline
our sample selection and observations in Section~\ref{sample_obs}.
We describe our analyses of these data in Section~\ref{stellar_prop}
and report the chemical abundances we infer in Section~\ref{chem_abund}.
We discuss the implications of our findings in Section~\ref{discussion}
and summarize our findings in Section~\ref{conclusion}.

\section{Sample Selection and Observations}\label{sample_obs}

As input to our candidate selection process we used the Two
Micron All Sky Survey (2MASS) All-Sky Point Source Catalog (PSC)
\citep{skrutskie2006} combined with Spitzer/IRAC Galactic Legacy
Infrared Mid-Plane Survey Extraordinaire (GLIMPSE) II and GLIMPSE
3D catalogs \citep{benjamin2003,churchwell2009}.  We dereddened and
extinction-corrected these photometric catalogs with the bulge-specific
reddening maps from \citet{gonzalez2011,gonzalez2012} assuming a
\citet{nishiyama2009} extinction law.  We excluded all stars with
non-zero data quality flags (indicating a possible data quality issue),
neighbors within 2\arcsec, or extinction-corrected apparent 2MASS
$J$-band magnitudes $J_0 \geq 12.5$.  The latter cut ensured that we
focused our attention on giant stars on the near side of the bulge.
We then applied the infrared-only ``v1'' metal-poor star selection from
\citet{schlaufman2014} to the dereddened and extinction-corrected 2MASS
and Spitzer/IRAC photometry to generate our initial candidate list.
The cuts described above resulted in a sample of 10,915 candidate
metal-poor giants with $-10 \leq l \leq +10$ and $-5 \leq b \leq
+5$.  To help prioritize spectroscopic follow-up, we also estimated
extinction-corrected $I$-band magnitudes assuming a typical $(I - J)_0 =
0.8$ color for metal-poor giants bright enough to have $J_0 \lesssim 12.5$
at 8 kpc.

\begin{figure*}
\plotone{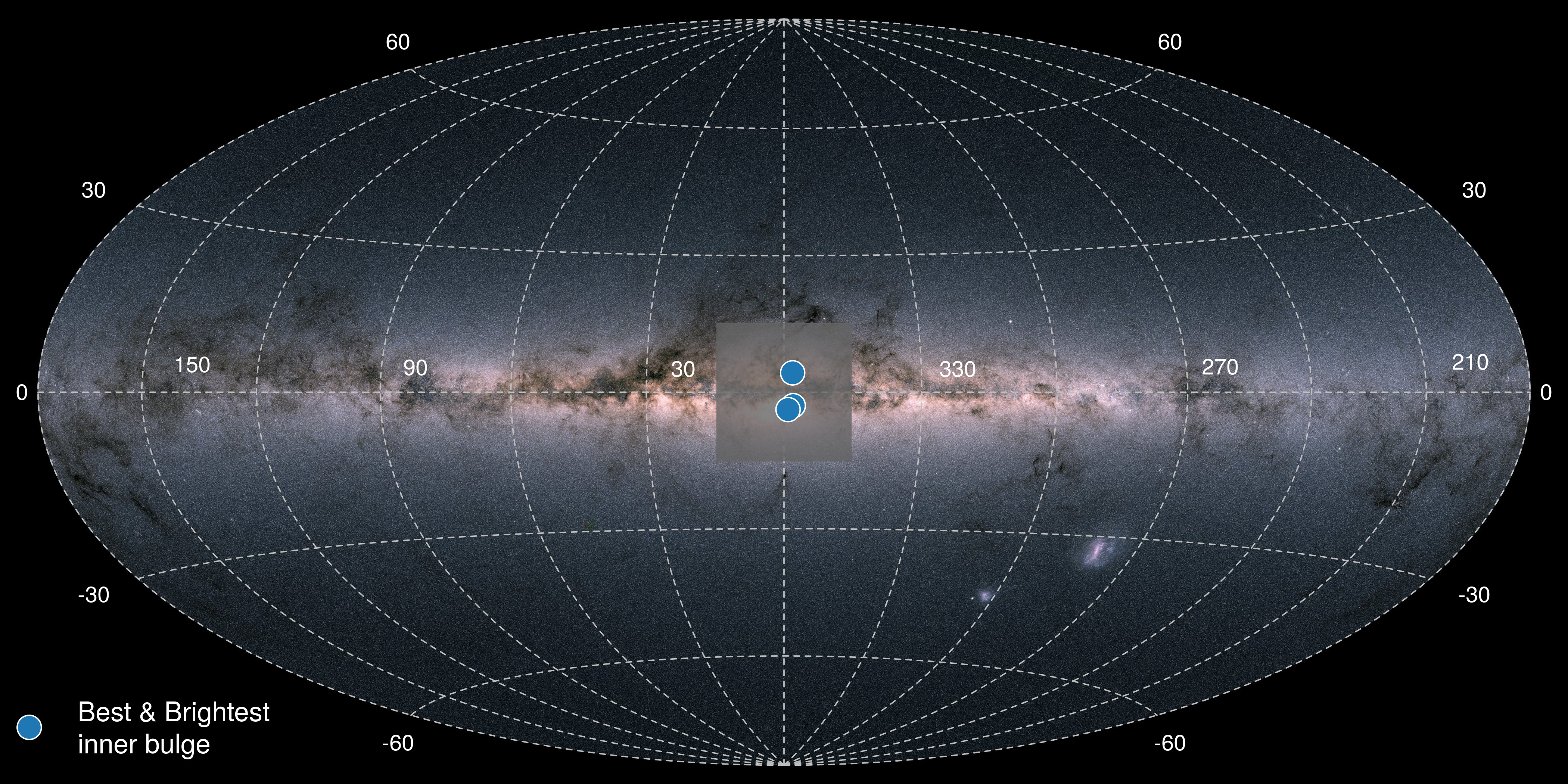}
\caption{Gaia DR2 image of the Milky Way. We indicate the locations
of the metal-poor inner bulge giants 2MASS J172452.74-281459.4, 2MASS
J175228.08-320947.6, and 2MASS J175836.79-313707.6 as blue points.
The shaded rectangle indicates the region in which \cite{tumlinson2010}
suggest that more than ten percent of stars with $[\mathrm{Fe/H}] \leq
-3$ formed at a time equivalent to $z > 15$, or more than 13.45 Gyr in
the past \citep{wright2006}.\label{stars_position}}
\end{figure*}

We observed several thousand of these candidate inner bulge metal-poor
giants using the Anglo-Australian Telescope's (AAT) AAOmega multiobject
spectrograph fed by the 2 Degree Field (2dF) robotic fibre positioner.
We used the 580V and 1700D gratings in the blue and red arms of the
spectrograph, providing spectral resolution $R \approx 1,\!300$ between
370 and 580 nm in the blue and $R \approx 10,\!000$ between 845 and 900
nm in the red.  We reduced these data using the standard \texttt{2dfdr}
pipeline.\footnote{\url{https://www.aao.gov.au/science/software/2dfdr}}
We estimated spectroscopic stellar parameters effective
temperature $T_{\mathrm{eff}}$, surface gravity $\log{g}$, and
metallicity $[\mathrm{Fe/H}]$ using the \texttt{sick} package
\citep{casey2016}.\footnote{\url{https://github.com/andycasey/sick}}
We then selected the giants 2MASS J172452.74-281459.4, 2MASS
J175228.08-320947.6, and 2MASS J175836.79-313707.6 for high-resolution
follow-up based on their low \texttt{sick}-inferred metallicities and
bright apparent magnitudes.  We plot the locations of these three stars on
the Gaia DR2 all-sky image of the Galaxy in Figure \ref{stars_position}.

We followed up these three giants with the Magellan Inamori Kyocera
Echelle (MIKE) spectrograph on the Magellan Clay Telescope at Las Campanas
Observatory \citep{bernstein2003,shectman2003}.  We used either the
0\farcs7~or 1\farcs0~slits and the standard blue and red grating azimuths,
yielding spectra between 335 nm and 950 nm with resolution $R \approx
40,\!000/28,\!000$ in the blue and $ R \approx 31,\!000/22,\!000$
in the red for the 0\farcs7/1\farcs0~slits.  We collected all
calibration data (e.g., bias, quartz \& ``milky" flat field, and ThAr
lamp frames) in the afternoon before each night of observations.
We present a log of these observations in Table~\ref{obs_log}.
We reduced the raw spectra and calibration frames using the
\texttt{CarPy}\footnote{\url{http://code.obs.carnegiescience.edu/mike}}
software package \citep{kelson2000,kelson2003,kelson2014}.  We used
\texttt{iSpec}\footnote{\url{https://www.blancocuaresma.com/s/iSpec}}
\citep{blanco-cuaresma2014,blanco-cuaresma2019} to calculate radial
velocities and barycentric corrections and normalized individual orders
using \texttt{IRAF}\footnote{\url{https://iraf-community.github.io/}}
\citep{iraf1986,iraf1993}.

\begin{deluxetable*}{llllccD}
\tablecaption{Log of Magellan/MIKE Observations\label{obs_log}}
\tablewidth{0pt}
\tablehead{
\colhead{Star} & \colhead{UT Date} & \colhead{Start} & \colhead{End} & \colhead{Slit Width} & \colhead{Exposure Time} & \twocolhead{RV} \\
\colhead{} & \colhead{} & \colhead{} & \colhead{} & \colhead{} & \colhead{(s)} & \twocolhead{(km s$^{-1}$)}}
\decimals
\startdata
J172452.74-281459.4 & 06/29/2017 & 04:21:29 & 04:32:29 & 1\farcs0 & 660 & +9.23 \\
J175228.08-320947.6 & 06/29/2017 & 04:34:37 & 04:54:37 & 1\farcs0 & 1200 & -62.52 \\
J175228.08-320947.6 & 07/02/2017 & 03:52:14 & 04:38:38 & 0\farcs7 & 2700& -63.98 \\
J175836.79-313707.6 & 06/29/2017 & 02:56:51 & 04:19:21 & 1\farcs0 & 4800 & -196.42  \\
J175836.79-313707.6 & 06/29/2017 & 04:56:40 & 05:26:40 & 1\farcs0 & 1800 & -196.42
\enddata
\end{deluxetable*}


The extreme extinction and reddening towards the inner bulge strongly
affected the signal-to-noise ratio S/N of our spectra blueward of
600 nm.  At 400 nm near the \ion{Ca}{2} H and K lines, our spectra
have $\mathrm{S/N} \approx 3$/pixel.  At 520 nm near the \ion{Mg}{1}
$b$ triplet our spectra have $\mathrm{S/N} \approx 10$/pixel, while at
660 nm near H$\alpha$ our spectra have $\mathrm{S/N} \approx 60$/pixel.
Near the near-infrared \ion{Ca}{2} triplet at 850 nm, our spectra have
$\mathrm{S/N} \approx 120$/pixel.  We therefore focused our absorption
line measurements on the long wavelength portions of our spectra.

\section{Stellar Properties}\label{stellar_prop}

\subsection{Stellar Parameters}

\begin{figure*}
\plotone{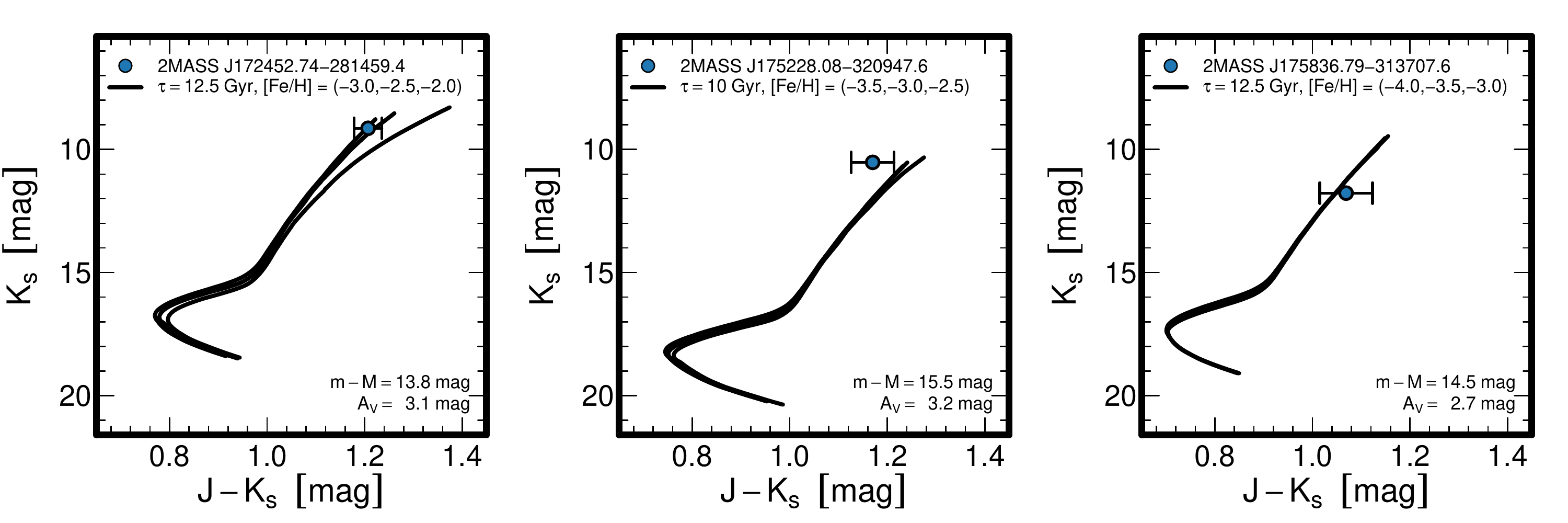}
\caption{Locations of the three stars in our sample relative to Dartmouth
isochrones with the indicated parameters.\label{iso}}
\end{figure*}

We used the
\texttt{isochrones}\footnote{\url{https://github.com/timothydmorton/isochrones}}
\citep{morton2015} package to estimate $T_{\mathrm{eff}}$ and $\log{g}$
of each star using as inputs their:
\begin{enumerate}
\item
$g$ and $r$ magnitudes and associated uncertainties from Data Release
(DR) 1.1 of the SkyMapper Southern Sky Survey \citep{wolf2018};
\item
$J$, $H$, and $K_{\mathrm{s}}$ magnitudes and associated uncertainties
from the 2MASS PSC \citep{skrutskie2006};
\item
$W1$, $W2$, and $W3$ magnitudes and associated uncertainties from
the Wide-field Infrared Survey Explorer (WISE) AllWISE Source Catalog
\citep{wright2010,mainzer2011};
\item
prior-informed distance estimates from
\citet{bailerjones2018} based on Gaia DR2 astrometry
\citep{gaia2016,gaia2018,arenou2018,hambly2018,lindegren2018,luri2018}.
\end{enumerate}
We used \texttt{isochrones} to fit the Dartmouth Stellar Evolution
Database \citep{dotter2007,dotter2008} library generated with the
Dartmouth Stellar Evolution Program (DSEP) to these observables using
\texttt{MultiNest}\footnote{\url{https://ccpforge.cse.rl.ac.uk/gf/project/multinest/}}
\citep{feroz2008,feroz2009,feroz2019}.  We restricted the Dartmouth
library to $\alpha$-enhanced composition $[\alpha\mathrm{/Fe}] = +0.4$,
stellar age $\tau$ in the range 10.0 Gyr $\leq \tau \leq$ 13.721 Gyr,
and extinction $A_{V}$ in the range 2.0 mag $\leq A_{V} \leq$ 5.0 mag.
For each star, we initially assumed the values $T_{\mathrm{eff}} =
4750 \pm 250$ K, $\log{g} = 2 \pm 1$, $[\mathrm{Fe/H}] = -3.0 \pm 1.0$
for the likelihood calculation.  We limited distances $d$ considered to
the range suggested by \citet{bailerjones2018}.  We plot the locations
of all three stars relative to isochrones in $J-K_{\mathrm{s}}$ versus
$K_{\mathrm{s}}$ color--magnitude diagrams in Figure \ref{iso} and give
the resulting isochrone-inferred parameters $T_{\mathrm{eff}}$, $\log{g}$,
$A_V$, $\tau$, stellar luminosity $L_{\ast}$, stellar mass $M_{\ast}$,
and isochrone distance $d_{\mathrm{iso}}$ in Table \ref{stellar_params}.
This approach is analogous to the \texttt{StarHorse} technique from
\citet{queiroz2018,queiroz2020}.

\begin{longrotatetable}
\begin{deluxetable*}{lcccl}
\tablecaption{Stellar Properties and Adopted Parameters\label{stellar_params}}
\tablewidth{0pt}
\tablehead{
\colhead{Property} & \colhead{J172452.74-281459.4} &
\colhead{J175228.08-320947.6} & \colhead{J175836.79-313707.6} &
\colhead{Units}}
\startdata
\textbf{Astrometric and Photometric Properties} & & & & \\
Gaia DR2 Source ID & 4059924905887808128 & 4043617705316006272 & 4043987927121580928 & \\
Gaia DR2 R.A. $\alpha$ (J2000) & 17 24 52.7503 & 17 52 28.0857 & 17 58 36.7952 & h m s \\
Gaia DR2 decl. $\delta$ (J2000) & -28 14 59.287 & -32 09 47.652 & -31 37 07.674 & d m s \\
Gaia DR2 galactic longitude $l$ (J2000) & 358.1186 & 357.9902 & 359.1173 & degrees \\
Gaia DR2 galactic latitude $b$ (J2000) & 4.2026 & -2.9241 & -3.7794 & degrees \\
Gaia DR2 proper motion (J2000) $\mu_{\alpha} \cos{\delta}$ & $-2.72 \pm 0.10$ & $-0.84 \pm 0.09$ & $-0.81 \pm 0.16$ & mas yr$^{-1}$ \\
Gaia DR2 proper motion (J2000) $\mu_{\delta}$ & $-10.28 \pm 0.07$ & $-3.12 \pm 0.08$ & $-7.65 \pm 0.13$ & mas yr$^{-1}$ \\
Gaia DR2 parallax $\pi$ (J2000) & $0.22 \pm 0.06$ & $0.04 \pm 0.06$ & $0.15 \pm 0.07$ & mas \\
SkyMapper $g$ & $15.560 \pm 0.010$ & $16.743 \pm 0.023$ & $17.261 \pm 0.027$ & AB mag \\
SkyMapper $r$ & $\cdots$ & $15.384 \pm 0.007$ & $15.967 \pm 0.007$ & AB mag \\
Gaia DR2 $G$ & $13.75 \pm 0.002$ & $14.95 \pm 0.002$ & $15.70 \pm 0.002$ & Vega mag \\
2MASS $J$ & $10.346 \pm 0.019$ & $11.694 \pm 0.024$ & $12.843 \pm 0.042$ & Vega mag \\
2MASS $H$ & $9.437 \pm 0.023$ & $10.788 \pm 0.025$ & $11.990 \pm 0.039$ & Vega mag \\
2MASS $K_{\mathrm{s}}$ & $9.139 \pm 0.021$ & $10.524 \pm 0.025$ & $11.774 \pm 0.034$ & Vega mag \\
WISE $W1$ & $8.936 \pm 0.023$ & $10.381 \pm 0.035$ & $\cdots$ & Vega mag \\
WISE $W2$ & $8.963 \pm 0.020$ & $10.450 \pm 0.036$ & $\cdots$ & Vega mag \\
WISE $W3$ & $8.995 \pm 0.041$ & $10.134 \pm 0.157$ & $\cdots$ & Vega mag \\
IRAC 3.6 & $8.915 \pm 0.042$ & $10.322 \pm 0.028$ & $11.647 \pm 0.040$ & Vega mag \\
IRAC 4.5 & $8.887 \pm 0.044$ & $10.309 \pm 0.029$ & $11.619 \pm 0.051$ & Vega mag \\
\hline
\textbf{Isochrone-inferred Parameters} & & & & \\
Effective temperature $T_{\mathrm{eff}}$ & $4360 \pm 10$ & $4760 \pm 10$ & $4900\pm 10$ & K \\
Surface gravity $\log{g}$ & $0.90\pm0.01$ & $0.84\pm0.01$ & $1.72_{-0.02}^{+0.03}$ & cm s$^{-2}$ \\
Luminosity $L_{\ast}$ & $900_{-20}^{+10}$ & $1570 \pm 10$ & $210 \pm 10$ & $L_{\odot}$ \\
Radius $R_{\ast}$ & $53 \pm 1$ & $62 \pm 1$ & $20 \pm 1$ & $R_{\odot}$ \\
Distance $d_{\mathrm{iso}}$ & $5.6 \pm 0.1$ & $12.4 \pm 0.1$ & $7.9_{-0.3}^{+0.2}$ & kpc \\
Mass $M_{\ast}$ & $0.80 \pm 0.01$ & $0.98 \pm 0.01$ & $0.79 \pm 0.01$ & $M_{\odot}$ \\
Age $\tau$ & $12.4 \pm 0.1$ & $10.0 \pm 0.1$ & $12.3 \pm 0.5$ & Gyr \\
Extinction $A_{V}$ & $3.10 \pm 0.02$ & $3.25 \pm 0.01$ & $2.71 \pm 0.03$ & mag \\
\hline
\textbf{Spectroscopy-inferred Properties and Parameters} & & & & \\
Radial velocity $v_r$ & $+9.23 \pm 1$ & $-63.98 \pm 1$ & $-196.42 \pm 1$ & km s$^{-1}$ \\
Microturbulence $\xi$ & $2.78^{+0.03}_{-0.01}$ & $3.22^{+0.05}_{-0.04}$ & $2.69 \pm 0.04$ & km s$^{-1}$ \\
Metallicity $[\mathrm{Fe/H}]$ & $-2.03 \pm 0.18$ & $-2.56 \pm 0.22$ & $-3.15 \pm 0.25$ & \\
Non-LTE corrected metallicity $[\mathrm{Fe/H}]_{\mathrm{NLTE}}$ & $-1.95$ & $-2.44$ & $-3.06$ & \\
\hline
\textbf{Galactic Orbit Parameters} & & & & \\
Total Galactic velocity $v$ & $89 \pm 3$ & $102 \pm 4$ & $220 \pm 3$ & km s$^{-1}$ \\
Pericenter of Galactic orbit $R_{\mathrm{peri}}$ & $0.3 \pm 0.1$ & $0.8 \pm 0.1$ & $0.1 \pm 0.1$ & kpc \\
Apocenter of Galactic orbit $R_{\mathrm{apo}}$ & $2.7 \pm 0.1$ & $4.5 \pm 0.1$ & $1.5_{-0.1}^{+0.2}$ & kpc \\
Eccentricity of Galactic orbit $e$ & $0.77_{-0.04}^{+0.03}$ & $0.70_{-0.03}^{+0.02}$ & $0.84_{-0.07}^{+0.03}$ & \\
Maximum distance from Galactic plane $z_{\mathrm{max}}$ & $1.18_{-0.04}^{+0.05}$ & $0.77_{-0.04}^{+0.03}$ & $0.92 \pm 0.05$ & kpc
\enddata
\end{deluxetable*}
\end{longrotatetable}

In parallel we obtained spectroscopic stellar parameter using the
classical excitation/ionization balance approach.  We measured
the equivalent widths of \ion{Fe}{1} and \ion{Fe}{2} atomic
absorption lines in our continuum-normalized spectra by fitting
Gaussian profiles with the \texttt{splot} task in \texttt{IRAF}.
We used the \texttt{deblend} task to disentangle absorption lines
from adjacent spectral features whenever necessary.  The atomic
data for \ion{Fe}{1} and \ion{Fe}{2} are from \texttt{linemake}
\citep{sneden2009,sneden2016} maintained by Vinicius Placco and
Ian Roederer\footnote{\url{https://github.com/vmplacco/linemake}}
as collected in Ji et al.\ (2020, submitted).  We report our input
atomic data, measured equivalent widths, and inferred abundances in
Table~\ref{measured_ews}.

We used 1D plane-parallel $\alpha$-enhanced ATLAS9 model atmospheres
\citep{castelli2004}, the 2019 version of the \texttt{MOOG}
radiative transfer code \citep{sneden1973}, and the \texttt{q$^2$}
\texttt{MOOG} wrapper\footnote{\url{https://github.com/astroChasqui/q2}}
\citep{ramirez2014} to calculate $T_{\mathrm{eff}}$, $\log{g}$,
$[\mathrm{Fe/H}]$, and microturbulence $\xi$ by simultaneously minimizing:
\begin{enumerate}
\item
the difference between our inferred \ion{Fe}{1} and \ion{Fe}{2}
abundances;
\item
the dependence of \ion{Fe}{1} abundance on excitation potential;
\item
the dependence of \ion{Fe}{1} abundance on reduced equivalent width.
\end{enumerate}
We initiated our optimization with reasonable guesses for
$[\mathrm{Fe/H}]$ and $\xi$ plus $T_{\mathrm{eff}}$ and
$\log{g}$ 500 K and 1.0 dex lower than the isochrone-inferred
parameters.  We find the spectroscopic stellar parameters listed in
Table~\ref{stellar_params_spec}.  Due to the low S/N of our spectra
blueward of 600 nm, we only analyzed \ion{Fe}{1} and \ion{Fe}{2} lines
with $\lambda \geq 500$ nm.  As a result, we cannot reliably measure
a large number of \ion{Fe}{1} lines over a wide range of excitation
potential.  In addition, most unblended \ion{Fe}{2} lines in the spectra
of metal-poor giants have $\lambda < 500$ nm.  These two issues make it
difficult to infer stellar parameters in a robust way using spectroscopy
alone.

\begin{deluxetable*}{lcccccc}
\tablecaption{Line List, Equivalent-width Measurements, and Abundances\label{measured_ews}}
\tablewidth{0pt}
\tablehead{
\colhead{Star} & \colhead{Wavelength} & \colhead{Species} &
\colhead{Excitation Potential} & \colhead{$\log{gf}$} &
\colhead{Equivalent Width} & \colhead{$\epsilon_{\mathrm{X}}$} \\
 & \colhead{(\AA)} & & \colhead{(eV)} & & (m\AA) & }
\startdata
J172452.74-281459.4 & $5682.633$ & \ion{Na}{1} & $2.102$ & $-0.706$ & $20.70$ & $4.330$\\ 
J172452.74-281459.4 & $5688.203$ & \ion{Na}{1} & $2.104$ & $-0.406$ & $36.50$ & $4.339$\\ 
J172452.74-281459.4 & $5889.951$ & \ion{Na}{1} & $0.000$ & $0.108$ & $344.00$ & $4.327$\\ 
J175228.08-320947.6 & $5682.633$ & \ion{Na}{1} & $2.102$ & $-0.706$ & $3.00$ & $3.728$\\ 
J175228.08-320947.6 & $5688.203$ & \ion{Na}{1} & $2.104$ & $-0.406$ & $2.50$ & $3.349$\\ 
J175836.79-313707.6 & $5889.951$ & \ion{Na}{1} & $0.000$ & $0.108$ & $141.00$ & $3.144$\\ 
J172452.74-281459.4 & $5528.405$ & \ion{Mg}{1} & $4.346$ & $-0.498$ & $124.00$ & $5.523$\\ 
J172452.74-281459.4 & $5711.088$ & \ion{Mg}{1} & $4.343$ & $-1.724$ & $51.20$ & $5.845$\\ 
J175228.08-320947.6 & $5172.684$ & \ion{Mg}{1} & $2.712$ & $-0.393$ & $245.40$ & $5.418$\\ 
J175228.08-320947.6 & $5183.604$ & \ion{Mg}{1} & $2.717$ & $-0.167$ & $213.00$ & $4.821$\\ 
J175228.08-320947.6 & $5528.405$ & \ion{Mg}{1} & $4.346$ & $-0.498$ & $132.80$ & $5.902$\\ 
J175228.08-320947.6 & $5711.088$ & \ion{Mg}{1} & $4.343$ & $-1.724$ & $16.20$ & $5.520$\\ 
J175836.79-313707.6 & $5172.684$ & \ion{Mg}{1} & $2.712$ & $-0.393$ & $160.60$ & $4.560$\\ 
J175836.79-313707.6 & $5183.604$ & \ion{Mg}{1} & $2.717$ & $-0.167$ & $181.70$ & $4.621$\\ 
\enddata
\tablecomments{This table is published in its entirety in the
machine-readable format.  A portion is shown here for guidance regarding
its form and content.}
\end{deluxetable*}

\begin{deluxetable*}{lRRRR}
\tablecaption{Spectroscopic Stellar Parameters\label{stellar_params_spec}}
\tablewidth{0pt}
\tablehead{
\colhead{Star} & \colhead{$T_{\mathrm{eff}}$} & \colhead{$\log{g}$} &
\colhead{$[\mathrm{Fe/H}]$} & \colhead{$\xi$} \\
\colhead{} & \colhead{(K)} & \colhead{} & \colhead{} & \colhead{(km s$^{-1}$)}}
\startdata
J172452.74-281459.4 & 4780 \pm 240 & 1.40 \pm 0.60 & -1.90 \pm 0.20 & 3.2 \pm 0.7 \\
J175228.08-320947.6 & 4790 \pm 210 & 0.47 \pm 0.51 & -2.57 \pm 0.20 & 3.4 \pm 0.9 \\
J175836.79-313707.6 & 4820 \pm 280 & 1.94 \pm 0.86 & -3.27 \pm 0.33 & 2.5 \pm 0.5
\enddata
\end{deluxetable*}

It has long been known that spectroscopic stellar parameters
inferred for metal-poor giants using the classical approach differ
from those derived using photometry and parallax information
\citep[e.g.,][]{korn2003,frebel2013,mucciarelli2020}.  Since local
thermodynamic equilibrium (LTE) is almost always assumed in the model
atmospheres used to interpret equivalent width measurements, these
differences are often attributed to the violation of the assumptions
of LTE in the photospheres of metal-poor giants.  As a result, we
impose the constraints on $T_{\mathrm{eff}}$ and $\log{g}$ deduced
from our isochrone analysis and use the same optimization strategy to
search for a self-consistent set of spectroscopic stellar parameters
$T_{\mathrm{eff}}$, $\log{g}$, $[\mathrm{Fe/H}]$, and $\xi$ using the
classical excitation/ionization balance approach.

We calculated our adopted [Fe/H] and $\xi$ uncertainties due to our
uncertain $T_{\mathrm{eff}}$ and $\log{g}$ estimates using a Monte Carlo
simulation.  On each iteration, we randomly sample self-consistent
pairs of $T_{\mathrm{eff}}$ and $\log{g}$ from our isochrone
posteriors and calculate the best [Fe/H] and $\xi$ using the classical
excitation/ionization balance approach.  After we find a converged
solution, we calculate mean iron abundances using our \ion{Fe}{1} and
\ion{Fe}{2} equivalent width measurements assuming the stellar parameters
found on that iteration.  We save the result of each iteration and
calculate [Fe/H] and its uncertainty as the (16,50,84) percentiles of
the resulting metallicity distribution ([Fe/H]$=-2.03^{+0.01}_{-0.01}$
for 2MASS J172452.74-281459.4, [Fe/H]$=-2.58^{+0.01}_{-0.01}$ for
2MASS J175228.08-320947.6, and [Fe/H]$=-3.15^{+0.02}_{-0.01}$ for 2MASS
J175836.79-313707.6).  We then take these uncertainties and the converged
stellar parameters described in the preceding paragraph and use them to
redo the isochrone calculation using the converged stellar parameters
and their uncertainties in the likelihood calculation.  We repeat this
entire process three times to obtain our final stellar parameters
presented in Table~\ref{stellar_params}. The final uncertainties
in Table~\ref{stellar_params} are larger than those described above
because the values in Table~\ref{stellar_params} account for both the
standard deviation in iron abundance inferred from individual lines and
the uncertainties due to our imperfectly estimated stellar parameters
derived from the Monte Carlo analysis.  The precise $T_{\mathrm{eff}}$
and $\log{g}$ resulting from our isochrone analysis imply that the
ultimate accuracy of our [Fe/H] estimate is limited by the uncertainties
in our measured equivalent widths. As a final check, we used our
measured \ion{Fe}{1} and \ion{Fe}{2} equivalent widths and initiated our
optimization process using the final set of stellar parameters listed in
Table~\ref{stellar_params}.  We find that the spectroscopically inferred
stellar parameters that result are consistent within their uncertainties
to our preferred values in Table~\ref{stellar_params}.

In addition, we used the \cite{casagrande2010} Infrared Flux
Method (IRFM) to verify our isochrone-inferred $T_{\mathrm{eff}}$.
We deredden the 2MASS $J-K_{\mathrm{s}}$ colors of our stars using the
bulge-specific reddening maps from \citet{gonzalez2011,gonzalez2012}.
In the IRFM calculation itself, we used the adopted $\log{g}$
and $[\mathrm{Fe/H}]$ given in Table~\ref{stellar_params}.
For 2MASS J172452.74-281459.4, 2MASS J175228.08-320947.6, and
2MASS J175836.79-313707.6 we find IRFM $T_{\mathrm{eff}} \approx
4530\pm350~\mathrm{K},4780\pm400~\mathrm{K},~\mathrm{and}~4680\pm400~\mathrm{K}$
in accord with the isochrone-inferred $T_{\mathrm{eff}}$ for each star.
The large uncertainties in the IRFM $T_{\mathrm{eff}}$ are due to the
reddening uncertainties in the \citet{gonzalez2011,gonzalez2012} map.

Finally, we calculated 1D non-LTE corrections to our individual iron
line abundances using the \cite{amarsi2016} grid.  While that grid
was calculated with MARCS model atmospheres \citep{gustafsson2008}
and our iron abundances were calculated with ATLAS9 model atmospheres,
both model atmospheres are very similar and we expect any differences
to have only a small effect on our abundance corrections.  We find that
the mean non-LTE corrections for \ion{Fe}{1} lines in our three giant
stars to be 0.18, 0.18, and 0.14 dex for 2MASS J172452.74-281459.4,
2MASS J175228.08-320947.6, and 2MASS J175836.79-313707.6.
The corrections are smaller for \ion{Fe}{2}: 0.03, 0.03, and 0.04
for 2MASS J172452.74-281459.4, 2MASS J175228.08-320947.6, and
2MASS J175836.79-313707.6.  These corrections are of the magnitude
expected for giant stars in this metallicity regime \citep[e.g.,
HD 122563 from][]{amarsi2016}.  We give our non-LTE [Fe/H] values in
Table~\ref{stellar_params}.

\subsection{Stellar Orbits}

To confirm that these giants located in the bulge are indeed on
tightly bound orbits, we calculated their Galactic orbits using
\texttt{galpy}\footnote{\url{https://github.com/jobovy/galpy}}.
We sampled 1,000 Monte Carlo realizations from the Gaia DR2 astrometric
solutions for each star using the distance posterior that results from
our isochrone analysis while taking full account of the covariances
between position, parallax, and proper motion.  We used the radial
velocities derived from our high-resolution MIKE spectra and assumed
no covariance between our measured radial velocity and the Gaia DR2
astrometric solution.  We used each Monte Carlo realization as an
initial condition for an orbit and integrated it forward 10 Gyr in
a Milky Way-like potential.  We adopted the \texttt{MWPotential2014}
described by \citet{bovy2015}.  In that model, the bulge is parameterized
as a power-law density profile that is exponentially cut-off at 1.9
kpc with a power-law exponent of $-1.8$.  The disk is represented by
a Miyamoto--Nagai potential with a radial scale length of 3 kpc and
a vertical scale height of 280 pc \citep{miyamoto1975}.  The halo is
modeled as a Navarro--Frenk--White halo with a scale length of 16 kpc
\citep{navarro1996}.  We set the solar distance to the Galactic center
to $R_{0} = 8.122$, kpc, the circular velocity at the Sun to $V_{0} =
238$ km s$^{-1}$, the height of the Sun above the plane to $z_{0} =
25$ pc, and the solar motion with the respect to the local standard
of rest to ($U_{\odot}$, $V_{\odot}$, $W_{\odot}$) = (10.0, 11.0,
7.0) km s$^{-1}$ \citep{juric2008,blandhawthorn2016,gravity2018}.
We give the resulting Galactic orbits in Table~\ref{stellar_params}.
We find that the Galactic orbits of all three giant stars have apocenters
$R_{\mathrm{apo}} \lesssim 4$ kpc, confirming that they are all indeed
tightly bound to the Galaxy and confined to the bulge region.

\section{Chemical Abundances}\label{chem_abund}

\begin{figure*}
\plotone{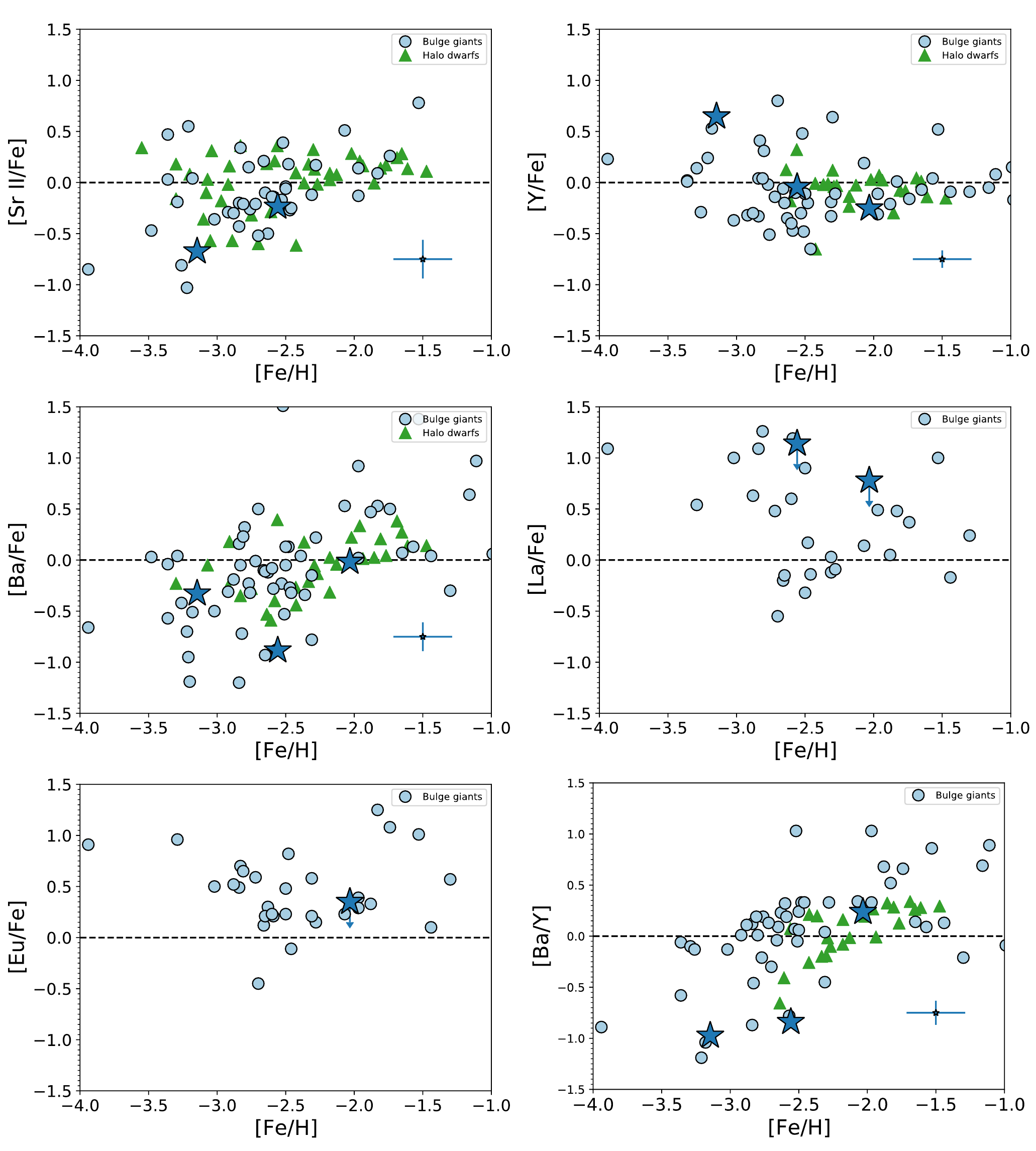}
\caption{Abundances of neutron-capture elements strontium, yttrium,
barium, lanthanum, and europium.  We plot as blue stars our three
metal-poor inner bulge giants.  We plot as light blue circles a literature
compilation of metal-poor outer bulge stars from \citet{garcia-perez2013},
\citet{casey2015}, \citet{howes2015,howes2016}, \citet{lamb2017},
and \citet{lucey2019}.  We plot as light green squares halo giants
from \cite{cayrel2004} and as dark green triangles halo dwarfs
from \cite{bonifacio2009} \& \cite{reggiani2017}.   The point with
error bars in the bottom right of each panel corresponds to the mean
uncertainty of our three stars.  We find that the neutron-capture
abundances of the inner and outer bulge are consistent with those in
the halo.\label{neutron_capture_fig}}
\end{figure*}

We measured the equivalent widths of atomic absorption lines for
\ion{Na}{1}, \ion{Mg}{1}, \ion{Al}{1}, \ion{Si}{1}, \ion{Ca}{1},
\ion{Sc}{2}, \ion{Ti}{1}, \ion{Ti}{2}, \ion{Cr}{1}, \ion{Cr}{2},
\ion{Mn}{1}, \ion{Co}{1}, \ion{Ni}{1}, \ion{Cu}{1}, \ion{Zn}{1},
\ion{Sr}{2}, \ion{Y}{2}, \ion{Ba}{2}, and \ion{La}{2} in our
continuum-normalized spectra by fitting Gaussian profiles with the
\texttt{splot} task in \texttt{IRAF}.  We used the \texttt{deblend}
task to disentangle absorption lines from adjacent spectral features
whenever necessary.  We measured an equivalent width for every
transition in our line list that could be recognized as an absorption
line regardless of S/N or wavelength, taking into consideration the
quality of a spectrum in the vicinity of a line and the availability
of alternative transitions of the same species.  We employed the 1D
plane-parallel $\alpha$-enhanced ATLAS9 model atmospheres and the 2019
version of \texttt{MOOG} to calculate abundances for each equivalent
width. In addition, we used spectral synthesis to infer the abundance
of \ion{Eu}{2} and to confirm the equivalent-width based abundance
of \ion{Ba}{2}.  We report our input atomic data from Ji et al.\ (2020,
submitted), measured equivalent widths, and individual inferred abundances
in Table~\ref{measured_ews}.  We present our adopted mean chemical
abundances and associated uncertainties in Table~\ref{chem_abundances}.
The standard deviation of abundances inferred for individual lines
$\sigma_{\epsilon}$ does not not take into account the uncertainties
in our adopted stellar parameters.  The uncertainties in individual
abundances relative to iron $\sigma_{[\mathrm{X/Fe}]}$ include both
the standard deviation of abundances inferred for individual lines
and the spectroscopic stellar parameter uncertainties assuming local
thermodynamic equilibrium.  We plot these later uncertainties in
Figures~\ref{alphas_fig}, \ref{light_odd_fig}, \ref{iron_peak_fig},
and \ref{neutron_capture_fig}.

\startlongtable
\begin{deluxetable*}{llrRrRRR}
\tabletypesize{\scriptsize}
\tablecaption{Chemical Abundances\label{chem_abundances}}
\tablecolumns{8}
\tablehead{
\colhead{Star} &
\colhead{Species} &
\colhead{N} &
\colhead{$\epsilon_{\mathrm{X}}$} &
\colhead{$\sigma_{\epsilon}$} &
\colhead{[X/H]} &
\colhead{[X/Fe]} &
\colhead{$\sigma_{[\mathrm{X/Fe}]}$}
}
\startdata
J172452.74-281459.4 & \ion{Na}{1} & $3$ & $4.332$ & $0.003$ & $-1.908$ & $0.124$ & $0.012$ \\ 
 & \ion{Na}{1}$_{\mathrm{NLTE}}$ & $3$ & $4.204$ & $\cdots$ & $-2.036$ & $-0.004$ & $\cdots$ \\ 
 & \ion{Mg}{1} & $2$ & $5.684$ & $0.114$ & $-1.916$ & $0.116$ & $0.161$ \\ 
 & \ion{Al}{1} & $2$ & $4.828$ & $0.066$ & $-1.622$ & $0.410$ & $0.093$ \\ 
 & \ion{Si}{1} & $5$ & $5.726$ & $0.026$ & $-1.784$ & $0.248$ & $0.029$ \\ 
 & \ion{Ca}{1} & $18$ & $4.607$ & $0.081$ & $-1.733$ & $0.299$ & $0.084$ \\ 
 & \ion{Sc}{2} & $5$ & $1.250$ & $0.073$ & $-1.900$ & $0.132$ & $0.082$ \\ 
 & \ion{Ti}{1} & $42$ & $3.157$ & $0.062$ & $-1.884$ & $0.148$ & $0.065$ \\ 
 & \ion{Ti}{2} & $50$ & $2.922$ & $0.126$ & $-2.028$ & $0.004$ & $0.127$ \\ 
 & $\overline{\mathrm{Ti}}$ & $\cdots$ & $3.029$ & $\cdots$ & $-1.921$ & $0.111$ & $0.106$ \\ 
 & \ion{Cr}{1} & $7$ & $3.254$ & $0.148$ & $-2.386$ & $-0.354$ & $0.161$ \\ 
 & \ion{Cr}{2} & $2$ & $4.027$ & $0.237$ & $-1.613$ & $0.419$ & $0.336$ \\ 
 & \ion{Mn}{1} & $3$ & $2.796$ & $0.014$ & $-2.634$ & $-0.602$ & $0.020$ \\ 
 & \ion{Fe}{1} & $22$ & $5.309$ & $0.174$ & $-2.191$ & $\cdots$ & $\cdots$ \\ 
 & \ion{Fe}{2} & $12$ & $5.626$ & $0.182$ & $-1.874$ & $\cdots$ & $\cdots$ \\ 
 & \ion{Co}{1} & $5$ & $3.146$ & $0.044$ & $-1.844$ & $0.188$ & $0.050$ \\ 
 & \ion{Ni}{1} & $14$ & $4.256$ & $0.056$ & $-1.964$ & $0.068$ & $0.059$ \\ 
 & \ion{Zn}{1} & $2$ & $2.958$ & $0.117$ & $-1.602$ & $0.430$ & $0.166$ \\ 
 & \ion{Sr}{1} & $1$ & $1.124$ & $0.000$ & $-1.746$ & $0.286$ & $0.016$ \\ 
 & \ion{Sr}{2} & $2$ & $\ge-1.280$ & $\cdots$ & $\ge-4.150$ & $\ge-2.118$ & $\cdots$ \\ 
 & \ion{Y}{2} & $3$ & $-0.076$ & $0.027$ & $-2.286$ & $-0.254$ & $0.034$ \\ 
 & \ion{Ba}{2} & $2$ & $0.131$ & $0.080$ & $-2.049$ & $-0.017$ & $0.113$ \\ 
 & \ion{La}{2} & $2$ & $\le-0.154$ & $\cdots$ & $\le-1.254$ & $\le0.778$ & $\cdots$ \\ 
J175228.08-320947.6 & \ion{Na}{1} & $2$ & $3.538$ & $0.134$ & $-2.702$ & $-0.144$ & $0.190$ \\ 
 & \ion{Na}{1}$_{\mathrm{NLTE}}$ & $2$ & $3.445$ & $\cdots$ & $-2.795$ & $-0.236$ & $\cdots$ \\ 
 & \ion{Mg}{1} & $4$ & $5.415$ & $0.194$ & $-2.185$ & $0.373$ & $0.225$ \\ 
 & \ion{Al}{1} & $1$ & $3.889$ & $0.000$ & $-2.561$ & $-0.003$ & $0.032$ \\ 
 & \ion{Si}{1} & $5$ & $5.646$ & $0.061$ & $-1.864$ & $0.694$ & $0.069$ \\ 
 & \ion{Ca}{1} & $15$ & $4.053$ & $0.071$ & $-2.287$ & $0.271$ & $0.073$ \\ 
 & \ion{Sc}{2} & $5$ & $0.665$ & $0.090$ & $-2.485$ & $0.073$ & $0.101$ \\ 
 & \ion{Ti}{1} & $28$ & $3.156$ & $0.119$ & $-1.885$ & $0.673$ & $0.121$ \\ 
 & \ion{Ti}{2} & $36$ & $2.414$ & $0.088$ & $-2.536$ & $0.022$ & $0.089$ \\ 
 & $\overline{\mathrm{Ti}}$ & $\cdots$ & $2.739$ & $\cdots$ & $-2.211$ & $0.347$ & $0.102$ \\ 
 & \ion{Cr}{1} & $5$ & $3.037$ & $0.190$ & $-2.603$ & $-0.045$ & $0.212$ \\ 
 & \ion{Cr}{2} & $4$ & $2.822$ & $0.109$ & $-2.818$ & $-0.260$ & $0.126$ \\ 
 & \ion{Mn}{1} & $3$ & $2.817$ & $0.138$ & $-2.613$ & $-0.055$ & $0.169$ \\ 
 & \ion{Fe}{1} & $50$ & $4.871$ & $0.218$ & $-2.629$ & $\cdots$ & $\cdots$ \\ 
 & \ion{Fe}{2} & $6$ & $5.055$ & $0.223$ & $-2.445$ & $\cdots$ & $\cdots$ \\ 
 & \ion{Co}{1} & $4$ & $3.199$ & $0.041$ & $-1.791$ & $0.767$ & $0.048$ \\ 
 & \ion{Ni}{1} & $9$ & $3.911$ & $0.074$ & $-2.309$ & $0.249$ & $0.079$ \\ 
 & \ion{Zn}{1} & $1$ & $2.251$ & $0.000$ & $-2.309$ & $0.249$ & $0.003$ \\ 
 & \ion{Sr}{1} & $1$ & $0.898$ & $0.000$ & $-1.972$ & $0.586$ & $0.007$ \\ 
 & \ion{Sr}{2} & $1$ & $0.076$ & $0.000$ & $-2.794$ & $-0.236$ & $0.040$ \\ 
 & \ion{Y}{2} & $2$ & $-0.392$ & $0.061$ & $-2.602$ & $-0.044$ & $0.087$ \\ 
 & \ion{Ba}{2} & $2$ & $-1.262$ & $0.077$ & $-3.442$ & $-0.884$ & $0.109$ \\ 
 & \ion{La}{2} & $2$ & $\le-0.318$ & $\cdots$ & $\le-1.418$ & $\le1.140$ & $\cdots$ \\ 
J175836.79-313707.6 & \ion{Na}{1} & $1$ & $3.144$ & $0.000$ & $-3.096$ & $0.050$ & $0.025$ \\ 
 & \ion{Na}{1}$_{\mathrm{NLTE}}$ & $1$ & $2.778$ & $\cdots$ & $-3.462$ & $-0.312$ & $\cdots$ \\ 
 & \ion{Mg}{1} & $2$ & $4.591$ & $0.022$ & $-3.009$ & $0.137$ & $0.041$ \\ 
 & \ion{Al}{1} & $1$ & $\le3.064$ & $\cdots$ & $\le-3.386$ & $\le-0.240$ & $\cdots$ \\ 
 & \ion{Si}{1} & $4$ & $5.050$ & $0.224$ & $-2.460$ & $0.686$ & $0.259$ \\ 
 & \ion{Ca}{1} & $15$ & $3.529$ & $0.088$ & $-2.811$ & $0.335$ & $0.092$ \\ 
 & \ion{Sc}{2} & $3$ & $0.390$ & $0.032$ & $-2.760$ & $0.386$ & $0.042$ \\ 
 & \ion{Ti}{1} & $27$ & $3.293$ & $0.115$ & $-1.748$ & $1.398$ & $0.119$ \\ 
 & \ion{Ti}{2} & $42$ & $2.489$ & $0.133$ & $-2.461$ & $0.685$ & $0.135$ \\ 
 & $\overline{\mathrm{Ti}}$ & $\cdots$ & $2.804$ & $\cdots$ & $-2.146$ & $1.000$ & $0.131$ \\ 
 & \ion{Cr}{1} & $4$ & $2.598$ & $0.014$ & $-3.042$ & $0.104$ & $0.025$ \\ 
 & \ion{Cr}{2} & $3$ & $2.926$ & $0.231$ & $-2.714$ & $0.432$ & $0.283$ \\ 
 & \ion{Mn}{1} & $3$ & $2.769$ & $0.067$ & $-2.661$ & $0.485$ & $0.083$ \\ 
 & \ion{Fe}{1} & $42$ & $4.325$ & $0.244$ & $-3.175$ & $\cdots$ & $\cdots$ \\ 
 & \ion{Fe}{2} & $6$ & $4.161$ & $0.264$ & $-3.339$ & $\cdots$ & $\cdots$ \\ 
 & \ion{Co}{1} & $1$ & $2.561$ & $0.000$ & $-2.429$ & $0.717$ & $0.020$ \\ 
 & \ion{Ni}{1} & $6$ & $3.442$ & $0.123$ & $-2.778$ & $0.368$ & $0.136$ \\ 
 & \ion{Cu}{1} & $1$ & $1.817$ & $0.000$ & $-2.373$ & $0.773$ & $0.020$ \\ 
 & \ion{Zn}{1} & $2$ & $2.209$ & $0.016$ & $-2.351$ & $0.795$ & $0.024$ \\ 
 & \ion{Sr}{2} & $2$ & $-0.950$ & $0.058$ & $-3.820$ & $-0.674$ & $0.086$ \\ 
 & \ion{Y}{2} & $3$ & $-0.289$ & $0.111$ & $-2.499$ & $0.647$ & $0.136$ \\ 
 & \ion{Ba}{2} & $3$ & $-1.293$ & $0.165$ & $-3.473$ & $-0.327$ & $0.202$ \\ 
 & \ion{La}{2} & $1$ & $\le-0.286$ & $\cdots$ & $\le-1.386$ & $\le1.760$ & $\cdots$
\enddata
\tablecomments{Abundance ratios assume \citet{asplund2009} solar
photospheric abundances.}
\end{deluxetable*}

To serve as comparison samples, we collected chemical abundances for
outer bulge stars and halo stars from the literature.  Our outer
bulge comparison sample comes from \cite{garcia-perez2013},
\citet{casey2015}, \citet{howes2015,howes2016}, \citet{lamb2017},
and \citet{lucey2019}.  Our halo comparison sample comes from
\citet{cayrel2004}, \citet{bonifacio2009}, and \citet{reggiani2017}.
We note that the current sample of metal-poor bulge stars shows
larger abundance dispersions than the sample of well-studied halo
stars for all elements.  These large dispersions are most likely due
to the lower S/N of the input bulge spectra combined with the lack of
a large-scale homogeneous abundance analyses in the metal-poor bulge.
On the other hand, it could also be that the metal-poor stars in the
bulge are first-generation Population II (Pop II) stars for which the
large dispersions appear because each star records the nucleosynthesis
of individual Population III (Pop III) supernovae.  While we regard the
former as more likely, only a large-scale homogeneous abundance analysis
of hundreds of metal-poor bulge stars will settle the issue.

\subsection{$\alpha$ Elements}\label{alpha_section}

Oxygen, magnesium, silicon, calcium, and titanium are often referred to
as $\alpha$ elements.  Magnesium, silicon, and calcium are formed via
similar nucleosynthetic channels.  Magnesium is mainly formed via carbon
burning in core-collapse supernovae (thermonuclear supernovae provide an
order of magnitude less).  Silicon is mostly a product of oxygen burning
and is itself the most abundant product of oxygen burning.  Core-collapse
and thermonuclear supernovae contribute to silicon production in equal
proportion.  Calcium is the product of both hydrostatic and explosive
oxygen and silicon burning.  It is mostly produced in core-collapse
supernovae.  Even though titanium forms either in the $\alpha$-rich
freeze-out of shock-decomposed nuclei during core-collapse supernovae
or in explosive $^{4}$He fusion in the envelopes of CO white dwarfs
during thermonuclear supernovae \citep[e.g.,][]{woosley1994,livne1995},
it is often considered alongside the true $\alpha$ elements because of
their correlated chemical abundances \citep{clayton2003}.

\begin{figure*}
\plotone{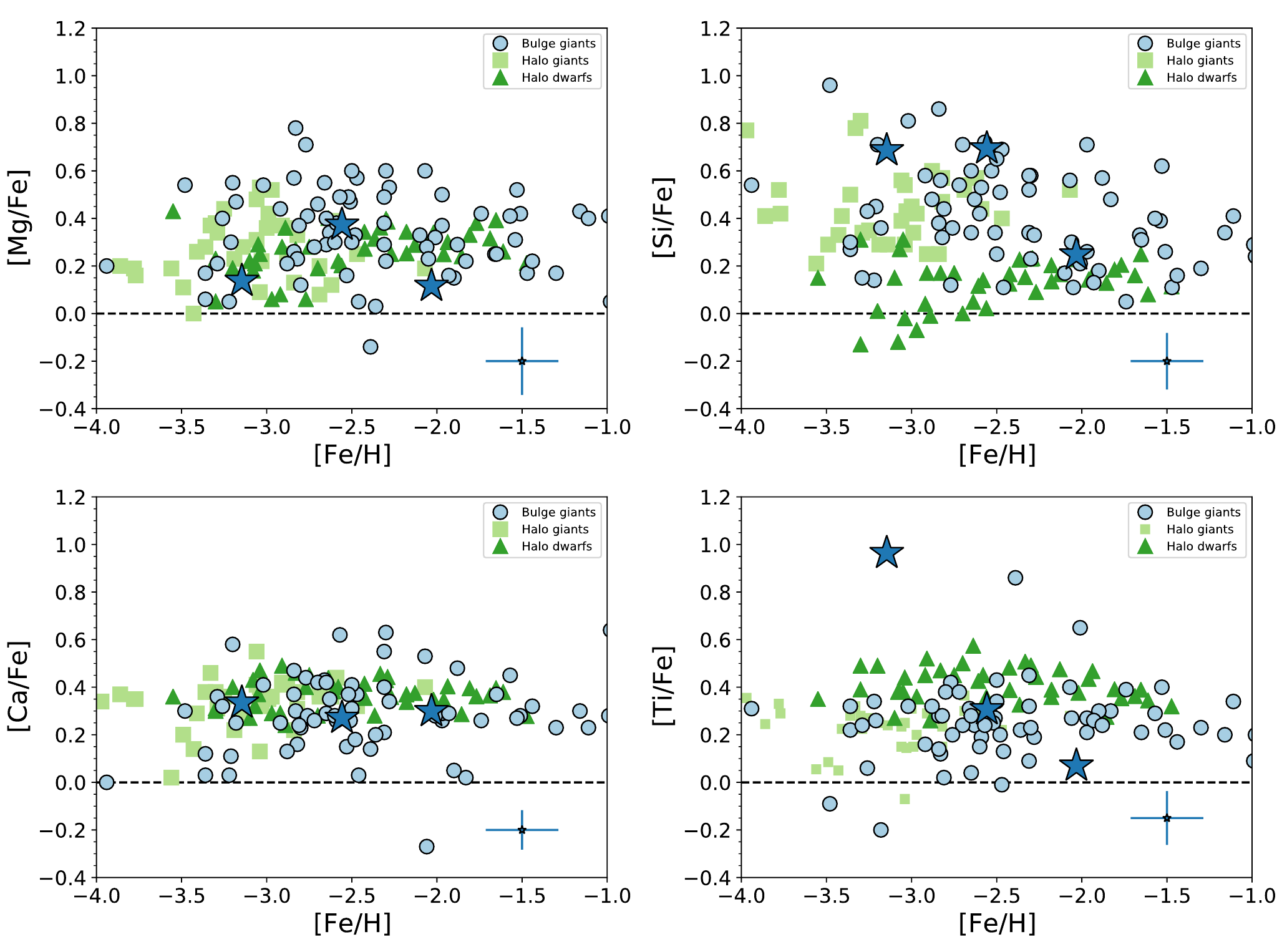}
\caption{Abundances of $\alpha$ elements magnesium, silicon, calcium,
and titanium relative to iron.  We plot as blue stars our three
metal-poor inner bulge giants.  We plot as light blue circles a literature
compilation of metal-poor outer bulge stars from \citet{garcia-perez2013},
\citet{casey2015}, \citet{howes2015,howes2016}, \citet{lamb2017},
and \citet{lucey2019}.  We plot as light green squares halo giants
from \cite{cayrel2004} and as dark green triangles halo dwarfs from
\cite{bonifacio2009} \& \cite{reggiani2017}.   The point with error bars
in the bottom right of each panel corresponds to the mean uncertainty of
our three stars.  We find that the $\alpha$-element abundances of the
inner and outer bulge are consistent with those in the halo.  The high
$[\mathrm{Si/Fe}] \approx+0.7$ abundance in our most metal-poor star 2MASS
J175836.79-313707.6 is suggestive of nucleosynthesis in an oxygen-rich
environment while the $[\mathrm{Ti/Fe}]\approx +1.0$ abundance could
be the result of a Chandrasekhar-mass thermonuclear supernova of a CO
white dwarf accreting from a helium star companion.\label{alphas_fig}}
\end{figure*}

\begin{figure}
\plotone{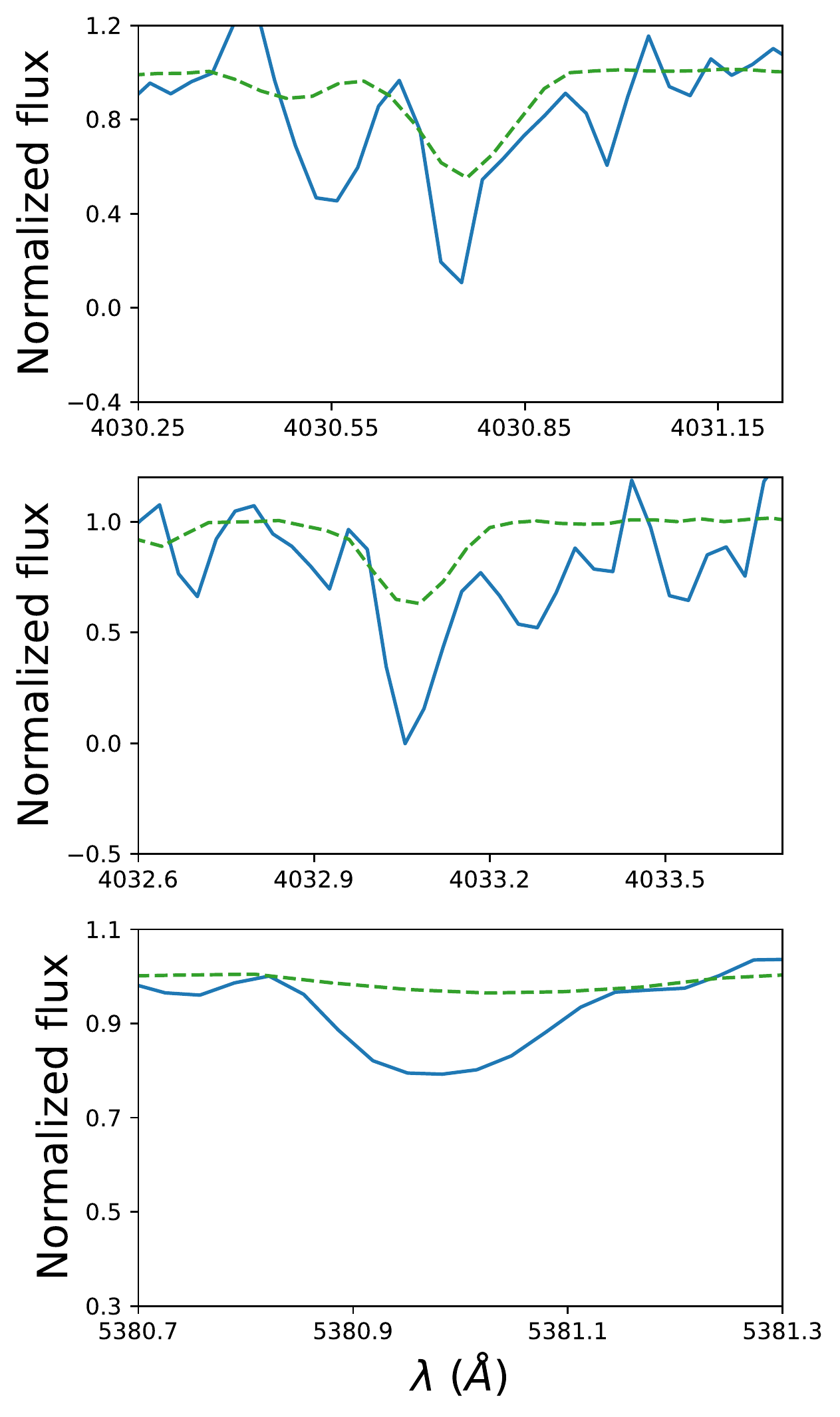}
\caption{Comparison of manganese and titanium lines for
2MASS J175836.79-313707.6 and BPS CS 30312-0059, a star from
\citet{roederer2014} with very similar spectroscopic stellar parameters
($T_{\mathrm{eff}} = 4780$ K, $\log{g} = 1.4$, and $[\mathrm{Fe/H}]
= -3.3$).  Top and middle: comparison of two \ion{Mn}{1} lines for
2MASS J175836.79-313707.6 (solid blue line) and BPS CS 30312-0059
(dashed green line)  Bottom: comparison of a \ion{Ti}{2} line for 2MASS
J175836.79-313707.6 (solid blue line) and BPS CS 30312-0059 (dashed
green line).\label{spec_lines}}
\end{figure}

We plot in Figure~\ref{alphas_fig} our inferred $\alpha$ abundances.
We find that our metal-poor giants in the inner bulge roughly
track the $\alpha$ abundances observed in the outer bulge and halo
comparison samples.  The one exception is the high \ion{Ti}{1} and
\ion{Ti}{2} abundances we infer for our most metal-poor star 2MASS
J175836.79-313707.6.  We plot in Figure~\ref{spec_lines} a representative
\ion{Ti}{2} line of 2MASS J175836.79-313707.6 in comparison to the same
line observed in BPS CS 30312-0059, a star from \citet{roederer2014}
with very similar spectroscopic stellar parameters ($T_{\mathrm{eff}} =
4780$ K, $\log{g} = 1.4$, and $[\mathrm{Fe/H}] = -3.3$).  We will argue
in Section~\ref{discussion} that the high titanium abundance in 2MASS
J175836.79-313707.6 is best explained by explosive $^{4}$He fusion in
the envelope of a CO white dwarf accreting from a helium star binary
companion during a Chandrasekhar-mass thermonuclear supernova.

The silicon abundances we infer from individual transitions for our three
inner bulge giants have non-negligible scatter and are affected by our
stellar parameter uncertainties.  Nevertheless, our inferred abundances
are in accord with silicon abundance inferences in outer bulge giants
with $[\mathrm{Fe/H}] \approx -3.2$ and follow the same trend observed
at higher metallicities.  We observe the largest silicon abundances
in our most metal-poor inner bulge giant 2MASS J175836.79-313707.6.
Most of the accessible silicon lines redward of 500 nm are weak in
such a metal-poor giant, so we also measured two additional lines
at 3906 and 4103 \AA~that were identifiable in its spectrum even at
$\mathrm{S/N} \approx 5$/pixel at 400 nm.  The apparent difference
between the silicon abundances inferred for halo dwarfs and giants is
usually attributed to non-LTE effects even though other factors play a
role \citep[e.g.,][Amarsi et al.\ 2020, submitted]{bonifacio2009b}.
According to the \citet{amarsi2017} grid of non-LTE
corrections\footnote{\url{http://www.mpia.de/homes/amarsi/index.html}},
the typical non-LTE silicon abundance correction for a giant star with
parameters similar to 2MASS J175836.79-313707.6 (i.e., $T_{\mathrm{eff}}
= 4500$ K, $\log{g} = 1.5$, $[\mathrm{Fe/H}] = -3.0$, $\xi = 2$ km
s$^{-1}$, and $\epsilon_{\mathrm{Si}}\approx 5.01$) is about $-0.02$
dex.  Non-LTE corrections are more important for metal-poor dwarfs,
as a dwarf with a similar metallicity ($T_{\mathrm{eff}} = 6500$ K,
$\log{g} = 4.5$, $[\mathrm{Fe/H}] = -3.0$, $\xi = 1$ km s$^{-1}$,
and $\epsilon_{\mathrm{Si}} \approx 5.01$) will have a non-LTE silicon
correction of about $+0.25$ dex.

\subsection{Light Odd-$Z$ Elements}\label{light_odd_section}

Like magnesium, sodium is mostly produced in core-collapse supernovae via
carbon burning.  Unlike magnesium, the surviving fraction of sodium in
supernovae ejecta depends on metallicity so it is treated as a secondary
product. Sodium is also produced as a product of hydrogen and helium
fusion in thermonuclear explosions, though in smaller quantities than
in core-collapse supernovae.  Similar to sodium, aluminum is synthesized
during carbon fusion in core-collapse supernovae in a secondary reaction
that is dependent on the amount of $^{22}$Ne burned (which in turn depends
on the carbon and oxygen content of the star). In contrast to sodium and
aluminum, scandium is formed via both oxygen burning in core-collapse
supernovae and as a product of $\alpha$-rich freeze-out in the shocked
region just above the rebounded core \citep[the same region responsible
for $^{44}$Ti e.g.,][]{clayton2003}.

\begin{figure*}
\plotone{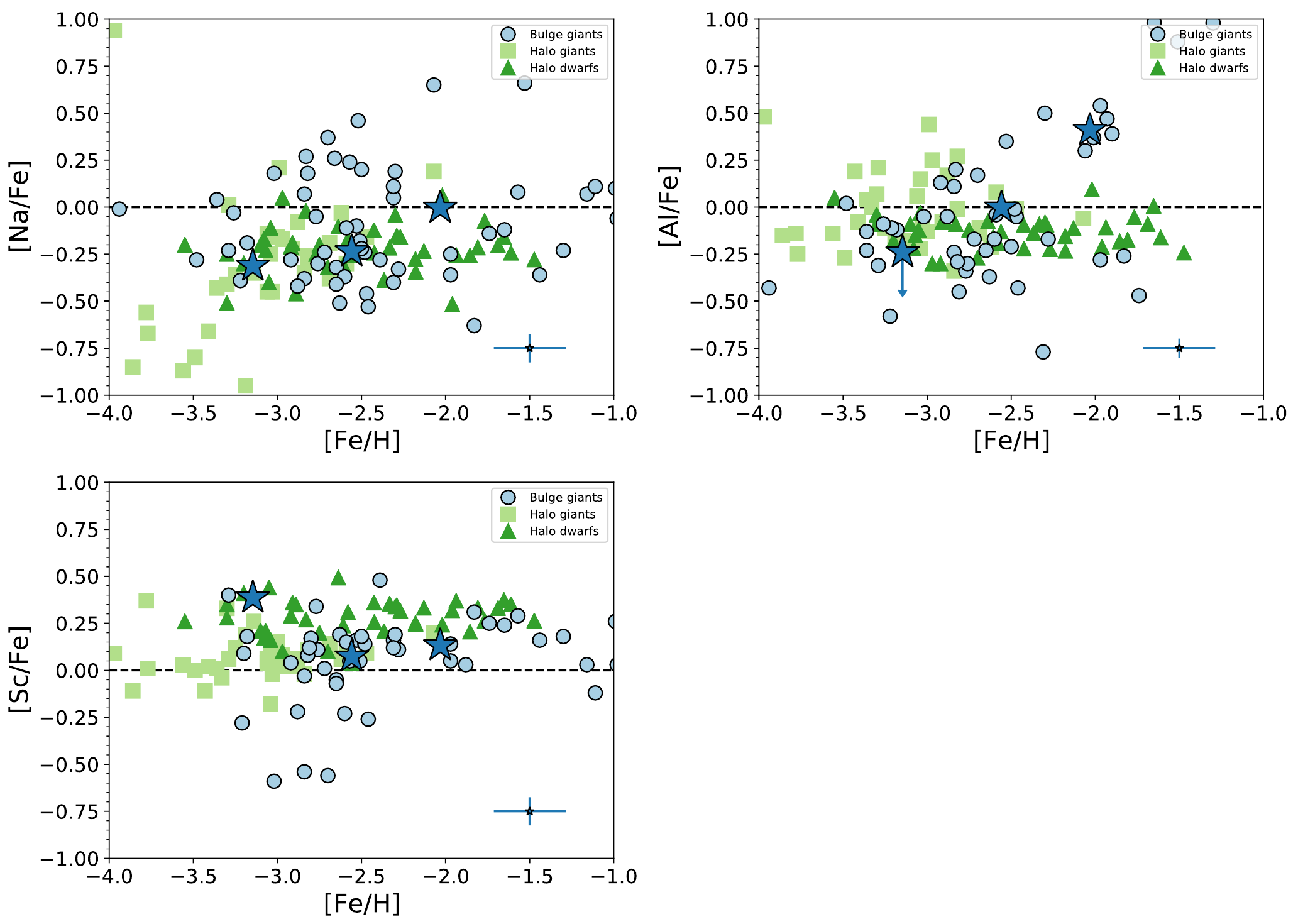}
\caption{Abundances of light odd-$Z$ elements sodium, aluminum, and
scandium relative to iron.  We plot as blue stars our three metal-poor
inner bulge giants.  We plot as light blue circles a literature
compilation of metal-poor outer bulge stars from \citet{garcia-perez2013},
\citet{casey2015}, \citet{howes2015,howes2016}, \citet{lamb2017},
and \citet{lucey2019}.  We plot as light green squares halo giants
from \cite{cayrel2004} and as dark green triangles halo dwarfs from
\cite{bonifacio2009} \& \cite{reggiani2017}.   The point with error bars
in the bottom right of each panel corresponds to the mean uncertainty
of our three stars.  We find that the light odd-$Z$ element abundances
of the inner and outer bulge are consistent with those in the halo.
The high $[\mathrm{Sc/Fe}] \approx +0.4$ in our most metal-poor star 2MASS
J175836.79-313707.6 is suggestive of nucleosynthesis in an oxygen-rich
environment.\label{light_odd_fig}}
\end{figure*}

We plot in Figure~\ref{light_odd_fig} our inferred light odd-$Z$
abundances.  We find that our metal-poor giants in the inner bulge roughly
track the light odd-$Z$ abundances observed in the outer bulge and halo
comparison samples.  The one exception is the high scandium abundance we
infer for our most metal-poor star 2MASS J175836.79-313707.6.  We will
argue in Section~\ref{discussion} that the high scandium abundance in
2MASS J175836.79-313707.6 is produced by nucleosynthesis in oxygen-rich
extreme Pop II stars.

Sodium abundance inferences are strongly affected by departures
from LTE, as the main sodium abundance indicator in our spectra
is the resonant sodium doublet at 5889/5895 \AA.  We corrected
our abundances inferred under the assumptions of LTE using
the \cite{lind2011} correction grid provided via the INSPECT
project\footnote{\url{http://inspect.coolstars19.com/}}.  To correct the
abundances of 2MASS J172452.74-281459.4 and 2MASS J175228.08-320947.6,
we used the correction for a star with $\log{g} = 1$ as our adopted
gravities were outside the bounds of the available grid.  The sodium
doublet is affected by interstellar medium (ISM) absorption, and the
extreme extinction along the line of sight to the inner bulge can
affect both the shape and depth of the sodium doublet.  As a result,
we were unable to disentangle the effects of photospheric and ISM
absorption for the 5895 \AA~line and we do not use it in our analysis.
Two weaker sodium lines at 5682 and 5688 \AA~are available in the
spectra of 2MASS J172452.74-281459.4 and 2MASS J175228.08-320947.6
though, and all abundances inferred from measured lines are in good
agreement.  For 2MASS J175836.79-313707.6, we only have the 5889 \AA~line.
Its spectrum has good S/N and is clear of ISM absorption, so we believe
our inferred sodium abundance is reliable.  Most of the sodium abundances
in our comparison outer bulge and halo samples have been corrected for
departures from local thermodynamic equilibrium.  \cite{cayrel2004} used
corrections from \citet{baumuller1998} while \cite{bonifacio2009} used
corrections from \cite{andrievsky2007}.  Like our sodium abundances,
\cite{howes2015,howes2016} and \cite{reggiani2017} were non-LTE
corrected using the grid from \citet{lind2011}.  \cite{casey2015} and
\cite{lucey2019} did not account for non-LTE effects.

It was extremely difficult to infer aluminum abundances for our three
stars.  The best available aluminum lines in metal-poor stars are usually
the 3944 and 3961 \AA~lines, and the spectra of our highly extincted inner
bulge stars have very low S/N at $\lambda < 400$ nm.  We were unable to
measure the equivalent width of either line in 2MASS J172452.74-281459.4.
We were only able to measure upper limits for the equivalent widths of
the 3944 \AA~line in 2MASS J175836.79-313707.6 and the 3961 \AA~line in
2MASS J175228.08-320947.6.  While there are two weaker aluminum lines at
6696 and 6698 \AA, they were only able to provide an upper limit on the
aluminum abundance of 2MASS J172452.74-281459.4.  While we report aluminum
abundances assuming LTE in Table~\ref{chem_abundances}, to fairly compare
our aluminum abundances with the outer bulge and halo samples we follow
\citet{reggiani2017} and add 0.65 dex to the LTE abundances of our three
stars as well as the LTE abundances in the comparison samples.

We inferred the scandium abundances of our three inner
bulge giants using \ion{Sc}{2} lines accounting for
hyperfine structure (HFS) using data taken from the Kurucz
compilation\footnote{\url{http://kurucz.harvard.edu/linelists.html}}.
Similar to what we observed with titanium, the abundance of scandium in
2MASS J175836.79-313707.6 is enhanced relative to the comparison samples.

\subsection{Iron-peak Elements}\label{iron_peak_section}

Iron-peak elements can be formed directly or as a byproduct of explosive
silicon burning, either incomplete (chromium and manganese) or complete
(cobalt, nickel, and zinc).  Their nucleosynthesis mainly takes place in
thermonuclear supernovae \citep[e.g.,][]{clayton2003,grimmett2019}. The
observed increase in the abundance ratios [Co/Fe] and [Zn/Fe] with
decreasing [Fe/H] in metal-poor stars combined with the dependence
of cobalt and zinc yields on the explosion energies of core-collapse
supernovae also point to contributions from hypernovae events at
$[\mathrm{Fe/H}] \lesssim -3.0$ \citep[e.g.,][]{cayrel2004,reggiani2017}.

\begin{figure*}
\plotone{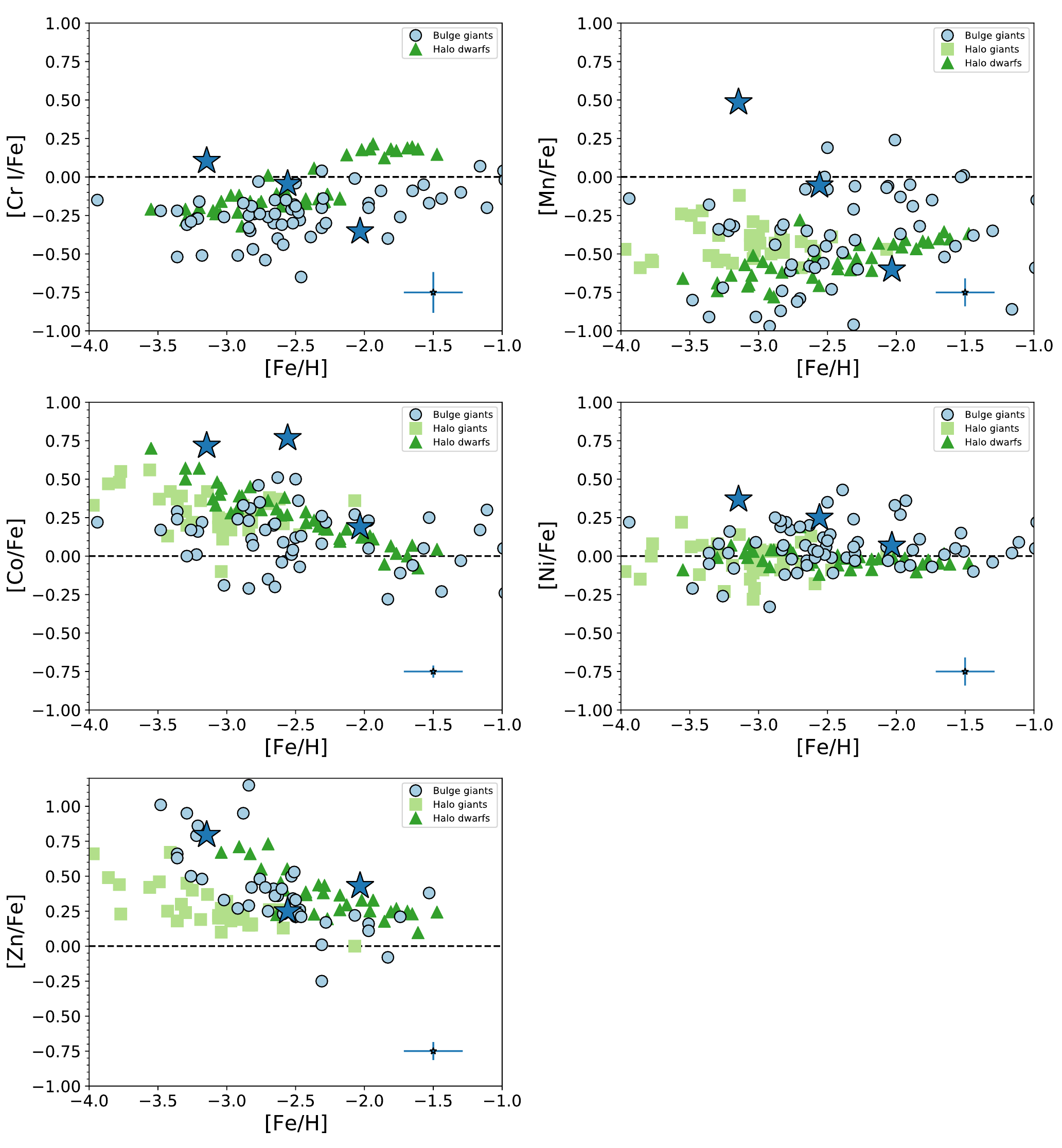}
\caption{Abundances of iron-peak elements chromium, manganese, cobalt,
nickel, and zinc.  We plot as blue stars our three metal-poor inner
bulge giants.  We plot as light blue circles a literature compilation
of metal-poor outer bulge stars from \citet{garcia-perez2013},
\citet{casey2015}, \citet{howes2015,howes2016}, \citet{lamb2017},
and \citet{lucey2019}.  We plot as light green squares halo giants
from \cite{cayrel2004} and as dark green triangles halo dwarfs from
\cite{bonifacio2009} \& \cite{reggiani2017}.   The point with error bars
in the bottom right of each panel corresponds to the mean uncertainty
of our three stars.  For our three inner bulge stars, we find iron-peak
abundances at the upper envelope of those observed in the outer bulge
and halo.  The significantly supersolar [Mn/Fe] abundance ratio we find
in our most metal-poor star 2MASS J175836.79-313707.6 is thought to be
a clear sign of nucleosynthesis in a Chandrasekhar-mass thermonuclear
supernova \citep[e.g.,][]{seitenzahl2013a}.\label{iron_peak_fig}}
\end{figure*}

We plot in Figure~\ref{iron_peak_fig} our inferred iron-peak abundances.
We find that our metal-poor giants in the inner bulge consistently have
higher iron-peak abundances than the outer bulge and halo comparison
samples.  We also find a significantly supersolar manganese abundance
$[\mathrm{Mn/Fe}] \approx +0.5$ for our most metal-poor star 2MASS
J175836.79-313707.6.  We do not correct our inferred iron-peak abundances
for non-LTE effects, both because correction grids are lacking for all
iron-peak elements and because the iron-peak abundances in our comparison
samples have not been corrected for departures from the assumptions
of LTE.  We will argue in Section~\ref{discussion} that the supersolar
manganese abundance in 2MASS J175836.79-313707.6---and indeed its entire
iron-peak abundance pattern---is best explained by nucleosynthesis in
a thermonuclear supernova.

We plot \ion{Cr}{1} abundances in Figure~\ref{iron_peak_fig} despite
the fact that those lines are strongly affected by departures from LTE
\citep[e.g.,][]{bergemann2010,reggiani2017}.  We prefer \ion{Cr}{1} to
\ion{Cr}{2} in this case because our inferred [\ion{Cr}{2}/Fe] ratio
in Table~\ref{chem_abundances} is significantly supersolar for 2MASS
J175836.79-313707.6.  Chromium almost always appears in solar [Cr/Fe]
ratios and neither core-collapse or thermonuclear supernovae produce
$[\mathrm{Cr/Fe}] \gtrsim +0.3$ \citep[e.g.,][]{clayton2003,grimmett2019}.
We therefore suspect that our inferred \ion{Cr}{2} abundances are affected
by noise in our spectra.

We find relatively high [Mn/Fe] abundances in our inner bulge
sample, including a significantly supersolar $[\mathrm{Mn/Fe}]
\approx +0.5$ in our most metal-poor star 2MASS J175836.79-313707.6.
We plot in Figure~\ref{spec_lines} three \ion{Mn}{1} lines for 2MASS
J175836.79-313707.6 in comparison to the same lines observed in the
comparison star BPS CS 30312-0059.  We included HFS components in our
abundance inferences using data taken from the Kurucz compilation
referenced above but did not correct for departures from LTE.
Corrections for departures from the assumptions of LTE in metal-poor
giants tend to increase [Mn/Fe] and would not change our conclusion about
2MASS J175836.79-313707.6 \citep[e.g.,][]{bergemann2019,eitner2020}.
The manganese abundance in 2MASS J175836.79-313707.6 is based on three
weak manganese lines at 6013, 6016, and 6021 \AA.  Our spectrum of
2MASS J175836.79-313707.6 has $\mathrm{S/N} \gtrsim 50$/pixel at 600
nm and even though they are at the limit of detectability, these three
manganese lines all appeared at their expected wavelengths and produce
a consistent manganese abundance estimate.  We therefore argue that the
apparent lines are unlikely to be produced by noise in our spectrum.
Although the bluer manganese lines typically analyzed in metal-poor
giants are apparent in the spectrum of 2MASS J175836.79-313707.6, the
abundances we infer from those lines are even higher.

\begin{figure}
\plotone{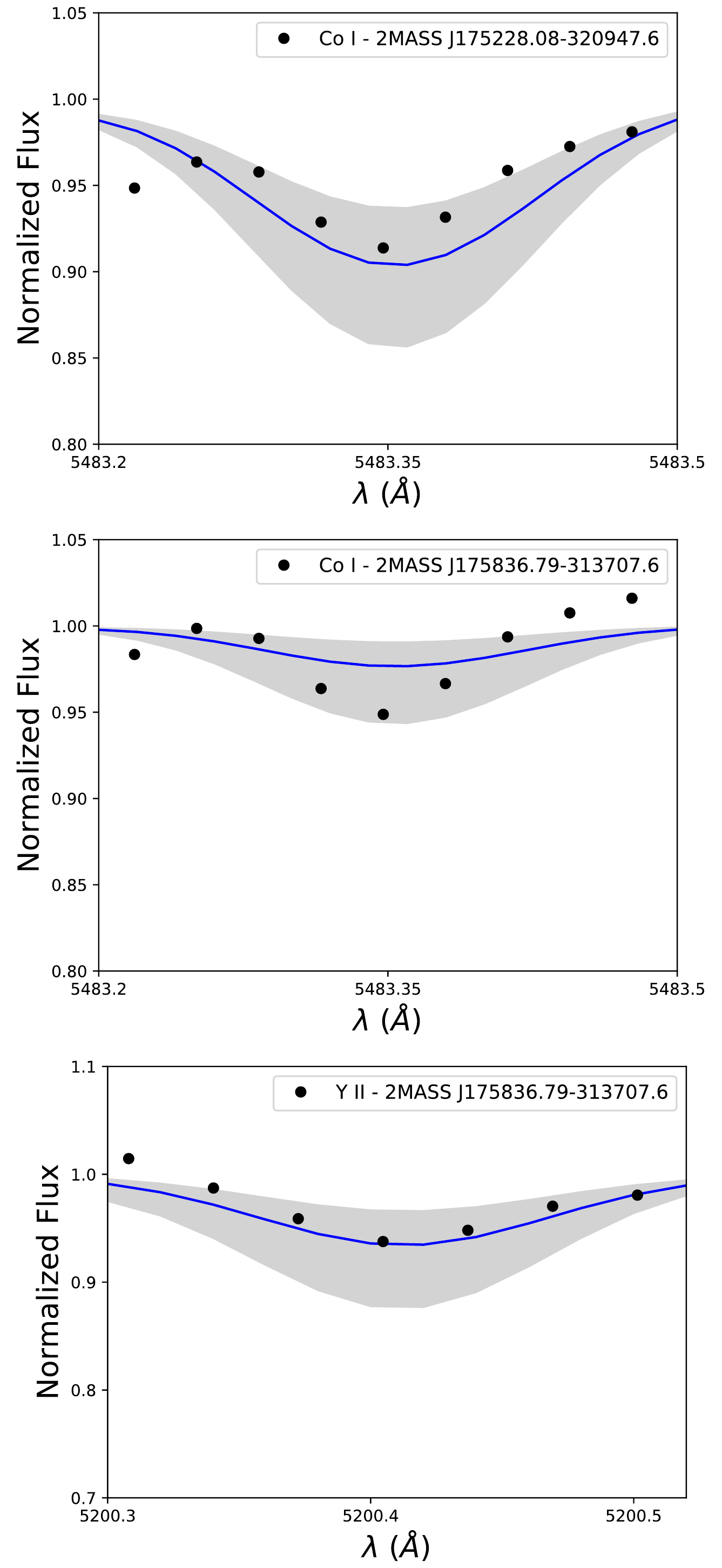}
\caption{Comparison between observed and synthesized lines of cobalt
and yttrium.  In each panel we plot the observed flux in the vicinity of
each line as black dots, the synthesized line assuming our adopted stellar
parameters and abundances in blue, and the uncertainty in the synthesized
line given our abundance uncertainties in gray.  Top: a \ion{Co}{1} line
in the star 2MASS J175228.08-320947.6 compared with the line synthesized
for cobalt abundance $\mathrm{A(Co)} = 3.20 \pm 0.3$.  Middle: the same
\ion{Co}{1} line in the star 2MASS J175836.79-313707.6 compared with the
line synthesized for cobalt abundance $\mathrm{A(Co)} = 2.56 \pm 0.3$.
Bottom: a \ion{Y}{2} line in the star 2MASS J175836.79-313707.6 compared
with the line synthesized for yttrium abundance $\mathrm{A(Y)}= -0.29
\pm 0.2$.\label{synth_lines}}
\end{figure}

We find relatively high cobalt abundances in our three metal-poor inner
bulge giants, especially in the range $-2.5 \lesssim [\mathrm{Fe/H}]
\lesssim -2.0$. A comparison of cobalt lines observed in the
spectra of 2MASS J175228.08-320947.6 and 2MASS J175836.79-313707.6 with
lines synthesized assuming our adopted stellar parameters and cobalt
abundances supports this finding (Figure~\ref{synth_lines}). At the
same time, our inferred zinc abundances closely track those observed in
the outer bulge and halo dwarf samples.  Our cobalt abundance estimates
come from lines redward of 500 nm, while our zinc abundance estimates are
based on lines blueward of 500 nm in the noisier parts of our spectra.
We find supersolar nickel abundances in our three inner bulge stars.
Like the outer bulge and halo comparison samples, we find no nickel
abundance trend with metallicity.  This lack of a dependence of nickel
abundance on metallicity supports the idea that nickel can be used as
a metallicity tracer in both the bulge and the halo \citep{singh2020}.

\subsection{Neutron-capture Elements}\label{neutron_capt_section}

Elements beyond zinc are mostly produced by neutron-capture processes
either ``slow'' or ``rapid'' relative to $\beta$ decay timescales.
The relative contributions of these $s$- and $r$-processes to
the nucleosynthesis of each element are different and functions
of metallicity.  Some elements like strontium, yttrium, barium, and
lanthanum are more commonly used as tracers of the $s$-process.  On the
other hand, europium is used as tracer of $r$-process nucleosynthesis
\citep[e.g.,][]{cescutti2006,jacobson2013,ji2016a}.  Even though strontium
and barium are usually used as tracers of the $s$-process, both can
have important contributions from $r$-process nucleosynthesis at lower
metallicities \citep[e.g.,][]{battistini2016,casey2017,mashonkina2019}.
Indeed, isotopic analyses of very metal-poor stars have shown that up
to 80\% of the barium in very metal-poor stars was synthesized in the
$r$-process \citep{mashonkina2019}.

We plot in Figure~\ref{iron_peak_fig} our inferred neutron-capture
element abundances.  We find that our metal-poor giants in the inner bulge
roughly track the neutron-capture abundances observed in the outer bulge
and halo comparison samples.  For strontium, we used \ion{Sr}{2} lines
to infer its elemental abundances as those lines are not significantly
affected by departures from LTE \citep{hansen2013}.  For yttrium we used
HFS components from the same Kurucz compilation referenced above, though
we note that the S/N of our spectra in the vicinity of the yttrium lines
were not high. Although we inferred our yttrium abundances using
equivalent widths, the observed spectrum of 2MASS J175836.79-313707.6
is in good agreement with a synthesized \ion{Y}{2} line at 5200
\AA~assuming our inferred abundance $\mathrm{A(Y)} = -0.29 \pm 0.2$
(Figure~\ref{synth_lines}).  For barium, the abundances we infer
using both equivalent widths and spectral synthesis based on unblended
\ion{Ba}{2} lines in parts of our spectra with high S/N agree within about
0.2 dex.  For europium and lanthanum, the low S/N of the blue parts of
our spectra only allow us to infer upper limits on their abundances.

\section{Discussion}\label{discussion}

The three metal-poor giants in the inner bulge we studied have orbits
that are confined to the bulge.  They are therefore likely to be among
the oldest stars in the Milky Way and trace the earliest stage of the
formation of the Milky Way's oldest component: the bulge.  We find that
the abundances of our inner bulge stars at $[\mathrm{Fe/H]} \gtrsim -3.0$
are for the most part in accord with the abundances of stars with similar
metallicities in the outer bulge and halo.

The story is different for our most metal-poor inner bulge giant 2MASS
J175836.79-313707.6 at $[\mathrm{Fe/H}] = -3.15$.  When compared to
both the outer bulge and halo comparison samples, it has high [Ti/Fe],
[Sc/Fe], and iron-peak abundances combined with supersolar [Mn/Fe].
We propose that 2MASS J175836.79-313707.6 is an ancient third-generation
star with $\alpha$ and light odd-$Z$ elements produced by massive Pop II
stars that were seeded with abundant oxygen by massive Pop III stars.
Unlike the progenitor(s) of the halo and the surviving dwarf galaxies,
the intense star formation rate in the bulge will fully sample the
stellar initial mass function and therefore produce many very massive
stars.  According to the Pop III supernovae yields of \citet{heger2010},
massive Pop III stars are prolific producers of oxygen relative to iron.
After their supernovae, that overabundance of oxygen is transformed into
an overabundance of scandium by the first generation of massive Pop II
stars and injected into the interstellar medium by their supernovae.
Titanium and the iron-peak elements including manganese in 2MASS
J175836.79-313707.6 were simultaneously produced in the Chandrasekhar-mass
thermonuclear supernova of a CO white dwarf accreting from a helium
star binary companion.  The combination of fast stellar evolution
at low metallicities, relatively massive CO white dwarfs produced by
metal-poor stars, and efficient accretion from a helium star produced a
short delay time comparable to the combined lifetimes of two generations
of massive stars, about 10 Myr after the onset of star formation in what
would become the bulge of the Milky Way.

To verify the scenario outlined above, we evaluate the ability
of models predicting the nucleosynthetic yields of core-collapse
and thermonuclear supernovae to reproduce the observed abundance
pattern of 2MASS J175836.79-313707.6.  Oxygen-rich metal-poor stars
produce more scandium relative to iron than solar-composition
metal-poor stars \citep[e.g.,][]{woosley1995,chieffi2004}.
The Chandrasekhar-mass thermonuclear supernova of a CO white
dwarf accreting from a helium star binary companion will produce
large amounts of titanium and can explode with a short delay time
\citep[e.g.,][]{woosley1994,livne1995,wang2009a,wang2009b}.  It is
thought that near Chandrasekhar-mass thermonuclear supernovae are the
only supernovae capable of producing $[\mathrm{Mn/Fe}] \gtrsim 0$,
as only CO white dwarfs near the Chandrasekhar mass have densities
$\rho \gtrsim 2\times10^{8}$ g cm$^{-3}$ necessary to produce
large amounts of $^{55}$Co that eventually decays into manganese
\citep[e.g.,][]{seitenzahl2013a,yamaguchi2015,seitenzahl2017}.

In an effort to confirm the scenario outlined above, we compare the
abundances of 2MASS J175836.79-313707.6 with the yields predicted by
three different classes of Chandrasekhar-mass thermonuclear supernovae
plus one class of sub-Chandrasekhar-mass thermonuclear supernovae:
\begin{enumerate}
\item
Chandrasekhar-mass deflagration to detonation transition (DDT) models
with fixed C/O ratios from \citet{seitenzahl2013b} and DDT models with
variable C/O ratios from \citet{ohlmann2014};
\item
Chandrasekhar-mass pure (turbulent) deflagration models from
\citet{fink2014};
\item
Chandrasekhar-mass gravitationally confined detonation (GCD) models
from \citet{seitenzahl2016};
\item
sub-Chandrasekhar-mass models resulting from the merger of two $M_{\ast}
= 0.6~M_{\odot}$ CO white dwarfs from \citet{papish2016}.
\end{enumerate}

We first fix the iron abundance predicted by each model to the metallicity
of 2MASS J175836.79-313707.6 and select the model that minimizes
$\chi^2$ between our observed abundances and the predicted yields.
If we focus only on the iron-peak abundances of chromium, manganese,
cobalt, and nickel we find that the pure deflagration model ``N100Hdef''
from \citet{fink2014} provides the best match to the abundances of 2MASS
J175836.79-313707.6.  The observable properties of model N100Hdef do not
match those of any known class of Type Ia supernovae though.  If instead
we consider both the abundances of silicon and the iron-peak elements
chromium, manganese, cobalt, and nickel we find that the DDT model
``N1600C'' from \citet{seitenzahl2013b} with a compact, spherically
symmetric ignition provides the best match to the abundances of 2MASS
J175836.79-313707.6.  In addition, model N1600C produces $[\mathrm{Si/Mg}]
\approx +0.9$ that is fully consistent with $[\mathrm{Si/Mg}] \approx
+0.6\pm0.3$ observed in 2MASS J175836.79-313707.6.  Moreover, the
observable properties of the DDT models from \citet{seitenzahl2013b}
do seem to match the properties of ordinary Type Ia supernovae.
The sub-Chandrasekhar-mass model is a poor fit to the abundances of
2MASS J175836.79-313707.6.

\begin{figure*}
\centering
\includegraphics[width=5in]{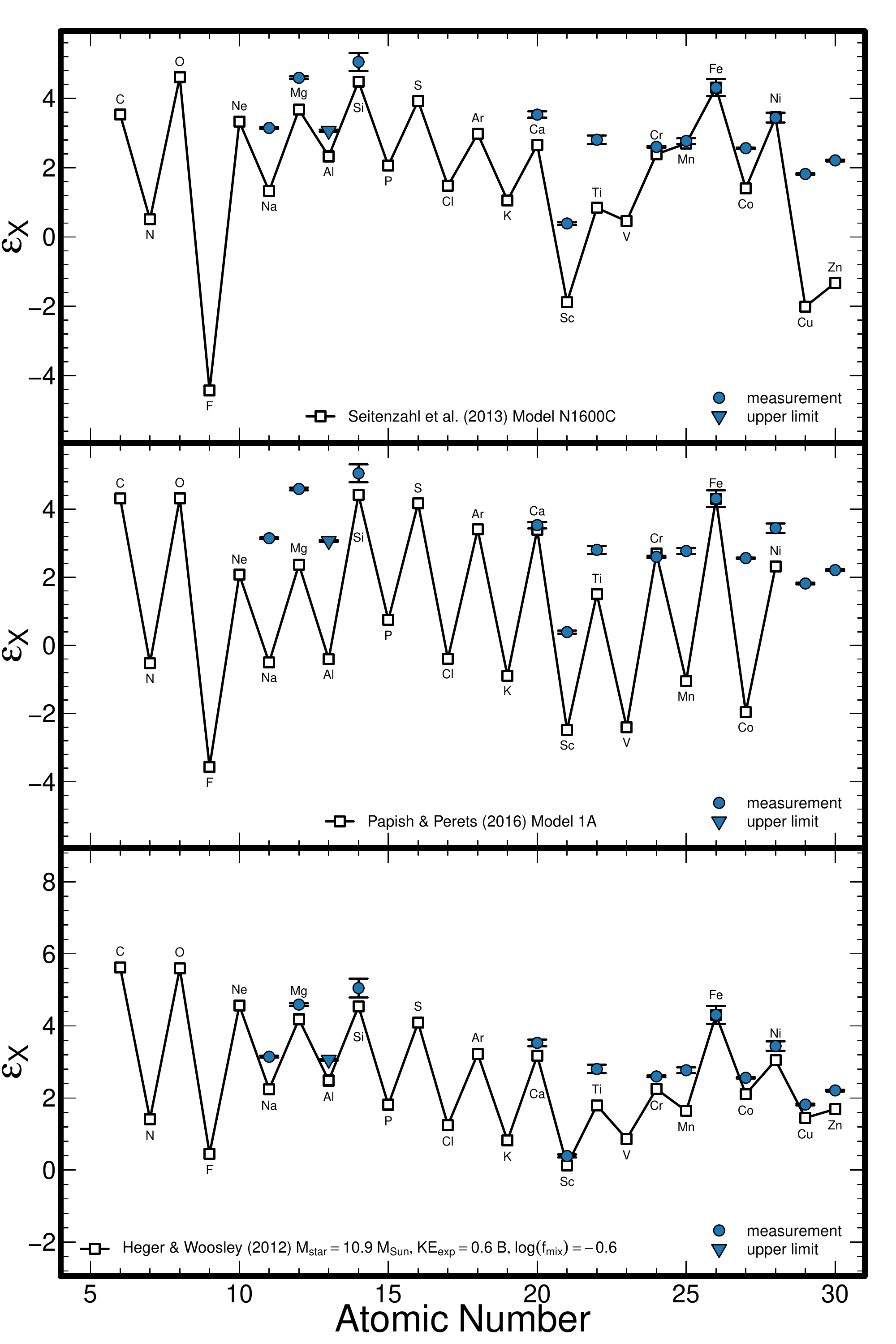}
\caption{Comparison of our inferred abundances for 2MASS
J175836.79-313707.6 with theoretical yields from supernovae models.
We minimize $\chi^2$ between our observed abundances and the
thermonuclear supernovae and core-collapse supernovae yields predicted
by \citet{seitenzahl2013b}, \citet{fink2014}, \citet{ohlmann2014},
\citet{seitenzahl2016}, \citet{papish2016}, and an updated version
of \cite{heger2010} after fixing the predicted abundances patterns to
match our inferred iron abundance.  Top: when considering both silicon
and the iron-peak elements chromium, manganese, cobalt, and nickel, we
find that the Chandrasekhar-mass deflagration to detonation transition
(DDT) model N1600C from \citet{seitenzahl2013b} best reproduces
our data.  Middle: the best-fit sub-Chandrasekhar-mass model ``1A''
from \citet{papish2016} is a poor fit to the abundances of 2MASS
J175836.79-313707.6. Bottom: the best-fit core-collapse supernova model
cannot reproduce the chromium, manganese, cobalt, or nickel abundances
of 2MASS J175836.79-313707.6.\label{thermo_model}}
\end{figure*}

We also compare the abundances of 2MASS J175836.79-313707.6 to the grid
of updated Pop III core-collapse supernovae ``znuc2012'' models from
\citet{heger2010} using STARFIT\footnote{\url{http://starfit.org/}}.
Though the best fit model ($M_{\ast}=10.9~M_{\odot}$, $KE_{\mathrm{exp}}
= 0.6$ B, $\log{f_{\mathrm{mix}}} = -0.6$) has a similar $\chi^2$
value as our preferred Chandrasekhar-mass thermonuclear model, it has
an implausibly large amount of mixing and cannot explain the chromium,
manganese, cobalt, or nickel abundances of 2MASS J175836.79-313707.6.
Indeed, the similar $\chi^2$ comes from the core-collapse model's
higher predicted cobalt yield that still fails to explain the cobalt
abundance of 2MASS J175836.79-313707.6.  Though the overall $\chi^2$
values are comparable, we prefer the Chandrasekhar-mass thermonuclear
supernovae model because it can self-consistently reproduce the chromium,
manganese, iron, and nickel abundances of 2MASS J175836.79-313707.6.
We plot in Figure~\ref{thermo_model} the best fit Chandrasekhar-mass
and sub-Chandrasekhar-mass thermonuclear supernovae models along with
the best-fit core-collapse supernova model.

Our scenario for the nucleosynthesis of the iron-peak elements in
2MASS J175836.79-313707.6 requires a Chandrasekhar-mass thermonuclear
supernovae to occur about 10 Myr after the formation of the first stars
in what would become the bulge of the Milky Way.  Chandrasekhar-mass
thermonuclear supernovae are often assumed to have occurred through
the so-called ``single degenerate'' channel in which a CO white dwarf
accretes material from Roche-lobe overflow or a strong wind from a
main sequence, subgiant, helium star, or red giant binary companion.
A single-degenerate Chandrasekhar-mass thermonuclear supernovae requires
a CO white dwarf.  At solar metallicity, the first CO white dwarfs
appear 30 to 40 Myr after the onset of star formation when stars less
massive than $8~M_{\odot}$ start to end their lives as white dwarfs.
Metal-poor stars both move through their stellar evolution more quickly
and produce more massive CO white dwarfs that require less accretion and
therefore less time to reach the Chandrasekhar mass than solar-metallicity
stars \citep[e.g.,][]{meng2008}.  The titanium abundance of 2MASS
J175836.79-313707.6 prefers accretion from a helium star companion,
and such configurations have been shown to reduce the delay times
of thermonuclear supernovae \citep[e.g.,][]{wang2009a,wang2009b}.
As a result, it seems plausible that Chandrasekhar-mass thermonuclear
supernovae delay times as short as 10 Myr might be achieved in a stellar
population with $[\mathrm{Fe/H}] \sim -3$.

Appealing to a thermonuclear supernova model that produces
$0.5~M_{\odot}$ of iron to explain the iron-peak abundances of a star
with $[\mathrm{Fe/H}] = -3.15$ requires either a significant outflow
of iron or an enormous dilution by unenriched gas.  The depth of the
potential at the center of the nascent Milky Way combined with the
presence of dense gas fueling ongoing star formation that would shock
ejecta and remove its kinetic energy indicates that dilution is a better
explanation.  To dilute $0.5~M_{\odot}$ of iron to the metallicity of
2MASS J175836.79-313707.6 requires mixing with about $10^{6}~M_{\odot}$
of pristine gas.  While this is a substantial amount of unenriched gas,
the young bulge is predicted to have had high accretions rates of pristine
gas due to the high gas densities and frequent mergers expected for a
relatively high $\sigma$ peak in the redshift $z\gtrsim2$ Universe.

The bulge of the young Milky Way is the ideal place to form stars
like 2MASS J175836.79-313707.6.  Even if the progenitor system of
a Chandrasekhar-mass thermonuclear supernova that explodes with a
delay time of 10 Myr is intrinsically rare, the star formation rate
is so high in the bulge of the young Milky Way that there are lots of
chances for it to form.  The high star formation rate will also fully
sample the stellar initial mass function and produce the massive Pop
III or extreme Pop II stars necessary to produce the oxygen that will
be transformed into silicon and scandium by future massive stars and
their supernovae.  The gas density and frequent mergers expected above $z
\approx 2$ will provide the unenriched gas necessary to form stars with
$[\mathrm{Fe/H}] \sim -3$ in a region with a high star formation rate.
We therefore suggest that stars with abundance patterns like that of
2MASS J175836.79-313707.6 enriched by thermonuclear supernovae should
only be observed in the halo or classical dwarf spheroidal galaxies at
higher metallicities.  Though uncommon, the abundance pattern of 2MASS
J175836.79-313707.6 is not unique.  \cite{reyes2020} inferred manganese
abundances for 161 giants in six classical dwarf spheroidal galaxies and
found [Mn/Fe] as high as our inferred value for 2MASS J175836.79-313707.6
in stars at approximately $-3.0 \lesssim [\mathrm{Fe/H}] \lesssim
-2.5$.  They concluded that their observed abundances are indicative
of Chandrasekhar-mass thermonuclear supernovae occurring at higher
metallicities $-2 \lesssim [\mathrm{Fe/H}] \lesssim -1$.

\section{Conclusion}\label{conclusion}

Metal-poor stars in the bulge on tightly bound orbits are thought to be
the oldest stars in the Milky Way.  We used the mid-infrared metal-poor
star selection of \citet{schlaufman2014} to find the three most metal-poor
stars known in the inner bulge.  All three stars are on tightly bound
orbits and confined to the bulge region.  The detailed abundances of
our two inner bulge giants with $[\mathrm{Fe/H}] \gtrsim -3$ have high
iron-peak abundances but are otherwise similar to metal-poor stars in the
outer bulge and halo.  Our most metal-poor star 2MASS J175836.79-313707.6
has high [Ti/Fe], [Sc/Fe], and iron-peak abundances.  It also has
supersolar [Mn/Fe].  We argue that it is a second-generation Pop II
star that was enriched by both massive Pop III or first-generation Pop
II stars and a Chandrasekhar-mass thermonuclear supernova accreting from
a helium star companion that exploded with a delay time of about 10 Myr.
We argue that stars like 2MASS J175836.79-313707.6 with $[\mathrm{Fe/H}]
\lesssim -3$ should be much more common in the bulge than in the halo
or dwarf galaxies because of the young bulge's high star formation rate
and frequent inflows of unenriched gas.

\acknowledgments
We thank the anonymous referee for the insightful comments that helped
us improve this paper. Andy Casey is the recipient of an Australian
Research Council Discovery Early Career Award (DECRA 190100656)
funded by the Australian Government.  Alex Ji is supported by NASA
through Hubble Fellowship grant HST-HF2-51393.001, awarded by the Space
Telescope Science Institute, which is operated by the Association of
Universities for Research in Astronomy, Inc., for NASA, under contract
NAS5-26555.  This work is based in part on observations made with
the Spitzer Space Telescope, which is operated by the Jet Propulsion
Laboratory, California Institute of Technology under a contract
with NASA.  This publication makes use of data products from the Two
Micron All Sky Survey, which is a joint project of the University of
Massachusetts and the Infrared Processing and Analysis Center/California
Institute of Technology, funded by the National Aeronautics and Space
Administration and the National Science Foundation.  This research has
made use of the NASA/IPAC Infrared Science Archive, which is funded
by the National Aeronautics and Space Administration and operated
by the California Institute of Technology.  Based in part on data
acquired through the Australian Astronomical Observatory, under programs
A/2015A/107 and A/2016A/103 plus Prop. ID 2016A-0086 and 2016B-0081 (PI:
K. Schlaufman).  We acknowledge the traditional owners of the land on
which the AAT stands, the Gamilaraay people, and pay our respects to
elders past and present.  This paper includes data gathered with the
6.5 m Magellan Telescopes located at Las Campanas Observatory, Chile.
This work has made use of data from the European Space Agency (ESA)
mission {\it Gaia} (\url{https://www.cosmos.esa.int/gaia}), processed
by the {\it Gaia} Data Processing and Analysis Consortium (DPAC,
\url{https://www.cosmos.esa.int/web/gaia/dpac/consortium}). Funding
for the DPAC has been provided by national institutions, in particular
the institutions participating in the {\it Gaia} Multilateral Agreement.
The national facility capability for SkyMapper has been funded through ARC
LIEF grant LE130100104 from the Australian Research Council, awarded to
the University of Sydney, the Australian National University, Swinburne
University of Technology, the University of Queensland, the University
of Western Australia, the University of Melbourne, Curtin University of
Technology, Monash University and the Australian Astronomical Observatory.
SkyMapper is owned and operated by The Australian National University's
Research School of Astronomy and Astrophysics. The survey data were
processed and provided by the SkyMapper Team at ANU.

\noindent
The SkyMapper node of the All-Sky Virtual Observatory (ASVO)
is hosted at the National Computational Infrastructure (NCI).
Development and support the SkyMapper node of the ASVO has been
funded in part by Astronomy Australia Limited (AAL) and the Australian
Government through the Commonwealth's Education Investment Fund (EIF)
and National Collaborative Research Infrastructure Strategy (NCRIS),
particularly the National eResearch Collaboration Tools and Resources
(NeCTAR) and the Australian National Data Service Projects (ANDS).
This publication makes use of data products from the \textit{Wide-field
Infrared Survey Explorer}, which is a joint project of the University of
California, Los Angeles, and the Jet Propulsion Laboratory/California
Institute of Technology, funded by the National Aeronautics and Space
Administration.  This research has made use of NASA's Astrophysics Data
System.  This research has made use of the SIMBAD database, operated
at CDS, Strasbourg, France \citep{wenger2000}.  This research has made
use of the VizieR catalogue access tool, CDS, Strasbourg, France (DOI:
10.26093/cds/vizier).  The original description of the VizieR service
was published in 2000, A\&AS 143, 23 \citep{ochsenbein2000}.

\vspace{5mm}
\facilities{AAT (2dF + AAOmega), CTIO:2MASS, Magellan:Clay (MIKE echelle
spectrograph), IRSA, Skymapper, Spitzer (IRAC), WISE.}

\software{\texttt{astropy} \citep{astropy2013,astropy2018},
          \texttt{CarPy} \citep{kelson2000,kelson2003},
          \texttt{galpy} \citep{bovy2015},
          \texttt{q2} \citep{ramirez2014},
          \texttt{isochrones} \citep{morton2015},
          \texttt{numpy} \citep{vanderwalt2011},
          \texttt{MultiNest} \citep{feroz2008,feroz2009,feroz2019},
          \texttt{pandas} \citep{scipy2010},
          \texttt{R} \citep{r20},
          \texttt{scipy} \citep{scipy2020}
          \texttt{IRAF} \citep{iraf1986,iraf1993}
          }

\bibliography{bulge}{}

\begin{thebibliography}{}
\expandafter\ifx\csname natexlab\endcsname\relax\def\natexlab#1{#1}\fi
\providecommand{\url}[1]{\href{#1}{#1}}
\providecommand{\dodoi}[1]{doi:~\href{http://doi.org/#1}{\nolinkurl{#1}}}
\providecommand{\doeprint}[1]{\href{http://ascl.net/#1}{\nolinkurl{http://ascl.net/#1}}}
\providecommand{\doarXiv}[1]{\href{https://arxiv.org/abs/#1}{\nolinkurl{https://arxiv.org/abs/#1}}}

\bibitem[{{Amarsi} \& {Asplund}(2017)}]{amarsi2017}
{Amarsi}, A.~M., \& {Asplund}, M. 2017, \mnras, 464, 264,
  \dodoi{10.1093/mnras/stw2445}

\bibitem[{{Amarsi} {et~al.}(2016){Amarsi}, {Lind}, {Asplund}, {Barklem}, \&
  {Collet}}]{amarsi2016}
{Amarsi}, A.~M., {Lind}, K., {Asplund}, M., {Barklem}, P.~S., \& {Collet}, R.
  2016, \mnras, 463, 1518, \dodoi{10.1093/mnras/stw2077}

\bibitem[{{Andrievsky} {et~al.}(2007){Andrievsky}, {Spite}, {Korotin}, {Spite},
  {Bonifacio}, {Cayrel}, {Hill}, \& {Fran{\c{c}}ois}}]{andrievsky2007}
{Andrievsky}, S.~M., {Spite}, M., {Korotin}, S.~A., {et~al.} 2007, \aap, 464,
  1081, \dodoi{10.1051/0004-6361:20066232}

\bibitem[{{Arenou} {et~al.}(2018){Arenou}, {Luri}, {Babusiaux}, {Fabricius},
  {Helmi}, {Muraveva}, {Robin}, {Spoto}, {Vallenari}, {Antoja},
  {Cantat-Gaudin}, {Jordi}, {Leclerc}, {Reyl{\'e}}, {Romero-G{\'o}mez}, {Shih},
  {Soria}, {Barache}, {Bossini}, {Bragaglia}, {Breddels}, {Fabrizio},
  {Lambert}, {Marrese}, {Massari}, {Moitinho}, {Robichon}, {Ruiz-Dern},
  {Sordo}, {Veljanoski}, {Eyer}, {Jasniewicz}, {Pancino}, {Soubiran}, {Spagna},
  {Tanga}, {Turon}, \& {Zurbach}}]{arenou2018}
{Arenou}, F., {Luri}, X., {Babusiaux}, C., {et~al.} 2018, \aap, 616, A17,
  \dodoi{10.1051/0004-6361/201833234}

\bibitem[{{Arentsen} {et~al.}(2020){Arentsen}, {Starkenburg}, {Martin}, {Hill},
  {Ibata}, {Kunder}, {Schultheis}, {Venn}, {Zucker}, {Aguado}, {Carlberg},
  {Gonz{\'a}lez Hern{\'a}ndez}, {Lardo}, {Longeard}, {Malhan}, {Navarro},
  {S{\'a}nchez-Janssen}, {Sestito}, {Thomas}, {Youakim}, {Lewis}, {Simpson}, \&
  {Wan}}]{arentsen2020}
{Arentsen}, A., {Starkenburg}, E., {Martin}, N.~F., {et~al.} 2020, \mnras, 491,
  L11, \dodoi{10.1093/mnrasl/slz156}

\bibitem[{{Asplund} {et~al.}(2009){Asplund}, {Grevesse}, {Sauval}, \&
  {Scott}}]{asplund2009}
{Asplund}, M., {Grevesse}, N., {Sauval}, A.~J., \& {Scott}, P. 2009, \araa, 47,
  481, \dodoi{10.1146/annurev.astro.46.060407.145222}

\bibitem[{{Astropy Collaboration} {et~al.}(2013){Astropy Collaboration},
  {Robitaille}, {Tollerud}, {Greenfield}, {Droettboom}, {Bray}, {Aldcroft},
  {Davis}, {Ginsburg}, {Price-Whelan}, {Kerzendorf}, {Conley}, {Crighton},
  {Barbary}, {Muna}, {Ferguson}, {Grollier}, {Parikh}, {Nair}, {Unther},
  {Deil}, {Woillez}, {Conseil}, {Kramer}, {Turner}, {Singer}, {Fox}, {Weaver},
  {Zabalza}, {Edwards}, {Azalee Bostroem}, {Burke}, {Casey}, {Crawford},
  {Dencheva}, {Ely}, {Jenness}, {Labrie}, {Lim}, {Pierfederici}, {Pontzen},
  {Ptak}, {Refsdal}, {Servillat}, \& {Streicher}}]{astropy2013}
{Astropy Collaboration}, {Robitaille}, T.~P., {Tollerud}, E.~J., {et~al.} 2013,
  \aap, 558, A33, \dodoi{10.1051/0004-6361/201322068}

\bibitem[{{Astropy Collaboration} {et~al.}(2018){Astropy Collaboration},
  {Price-Whelan}, {Sip{\H{o}}cz}, {G{\"u}nther}, {Lim}, {Crawford}, {Conseil},
  {Shupe}, {Craig}, {Dencheva}, {Ginsburg}, {Vand erPlas}, {Bradley},
  {P{\'e}rez-Su{\'a}rez}, {de Val-Borro}, {Aldcroft}, {Cruz}, {Robitaille},
  {Tollerud}, {Ardelean}, {Babej}, {Bach}, {Bachetti}, {Bakanov}, {Bamford},
  {Barentsen}, {Barmby}, {Baumbach}, {Berry}, {Biscani}, {Boquien}, {Bostroem},
  {Bouma}, {Brammer}, {Bray}, {Breytenbach}, {Buddelmeijer}, {Burke},
  {Calderone}, {Cano Rodr{\'\i}guez}, {Cara}, {Cardoso}, {Cheedella}, {Copin},
  {Corrales}, {Crichton}, {D'Avella}, {Deil}, {Depagne}, {Dietrich}, {Donath},
  {Droettboom}, {Earl}, {Erben}, {Fabbro}, {Ferreira}, {Finethy}, {Fox},
  {Garrison}, {Gibbons}, {Goldstein}, {Gommers}, {Greco}, {Greenfield},
  {Groener}, {Grollier}, {Hagen}, {Hirst}, {Homeier}, {Horton}, {Hosseinzadeh},
  {Hu}, {Hunkeler}, {Ivezi{\'c}}, {Jain}, {Jenness}, {Kanarek}, {Kendrew},
  {Kern}, {Kerzendorf}, {Khvalko}, {King}, {Kirkby}, {Kulkarni}, {Kumar},
  {Lee}, {Lenz}, {Littlefair}, {Ma}, {Macleod}, {Mastropietro}, {McCully},
  {Montagnac}, {Morris}, {Mueller}, {Mumford}, {Muna}, {Murphy}, {Nelson},
  {Nguyen}, {Ninan}, {N{\"o}the}, {Ogaz}, {Oh}, {Parejko}, {Parley}, {Pascual},
  {Patil}, {Patil}, {Plunkett}, {Prochaska}, {Rastogi}, {Reddy Janga},
  {Sabater}, {Sakurikar}, {Seifert}, {Sherbert}, {Sherwood-Taylor}, {Shih},
  {Sick}, {Silbiger}, {Singanamalla}, {Singer}, {Sladen}, {Sooley},
  {Sornarajah}, {Streicher}, {Teuben}, {Thomas}, {Tremblay}, {Turner},
  {Terr{\'o}n}, {van Kerkwijk}, {de la Vega}, {Watkins}, {Weaver}, {Whitmore},
  {Woillez}, {Zabalza}, \& {Astropy Contributors}}]{astropy2018}
{Astropy Collaboration}, {Price-Whelan}, A.~M., {Sip{\H{o}}cz}, B.~M., {et~al.}
  2018, \aj, 156, 123, \dodoi{10.3847/1538-3881/aabc4f}

\bibitem[{{Bailer-Jones} {et~al.}(2018){Bailer-Jones}, {Rybizki}, {Fouesneau},
  {Mantelet}, \& {Andrae}}]{bailerjones2018}
{Bailer-Jones}, C.~A.~L., {Rybizki}, J., {Fouesneau}, M., {Mantelet}, G., \&
  {Andrae}, R. 2018, \aj, 156, 58, \dodoi{10.3847/1538-3881/aacb21}

\bibitem[{{Barbuy} {et~al.}(2018){Barbuy}, {Chiappini}, \&
  {Gerhard}}]{barbuy2018}
{Barbuy}, B., {Chiappini}, C., \& {Gerhard}, O. 2018, \araa, 56, 223,
  \dodoi{10.1146/annurev-astro-081817-051826}

\bibitem[{{Battistini} \& {Bensby}(2016)}]{battistini2016}
{Battistini}, C., \& {Bensby}, T. 2016, \aap, 586, A49,
  \dodoi{10.1051/0004-6361/201527385}

\bibitem[{{Baumueller} {et~al.}(1998){Baumueller}, {Butler}, \&
  {Gehren}}]{baumuller1998}
{Baumueller}, D., {Butler}, K., \& {Gehren}, T. 1998, \aap, 338, 637

\bibitem[{{Benjamin} {et~al.}(2003){Benjamin}, {Churchwell}, {Babler}, {Bania},
  {Clemens}, {Cohen}, {Dickey}, {Indebetouw}, {Jackson}, {Kobulnicky},
  {Lazarian}, {Marston}, {Mathis}, {Meade}, {Seager}, {Stolovy}, {Watson},
  {Whitney}, {Wolff}, \& {Wolfire}}]{benjamin2003}
{Benjamin}, R.~A., {Churchwell}, E., {Babler}, B.~L., {et~al.} 2003, \pasp,
  115, 953, \dodoi{10.1086/376696}

\bibitem[{{Bensby} {et~al.}(2017){Bensby}, {Feltzing}, {Gould}, {Yee},
  {Johnson}, {Asplund}, {Mel{\'e}ndez}, {Lucatello}, {Howes}, {McWilliam},
  {Udalski}, {Szyma{\'n}ski}, {Soszy{\'n}ski}, {Poleski}, {Wyrzykowski},
  {Ulaczyk}, {Koz{\l}owski}, {Pietrukowicz}, {Skowron}, {Mr{\'o}z}, {Pawlak},
  {Abe}, {Asakura}, {Bhattacharya}, {Bond}, {Bennett}, {Hirao}, {Nagakane},
  {Koshimoto}, {Sumi}, {Suzuki}, \& {Tristram}}]{bensby2017}
{Bensby}, T., {Feltzing}, S., {Gould}, A., {et~al.} 2017, \aap, 605, A89,
  \dodoi{10.1051/0004-6361/201730560}

\bibitem[{{Bergemann} \& {Cescutti}(2010)}]{bergemann2010}
{Bergemann}, M., \& {Cescutti}, G. 2010, \aap, 522, A9,
  \dodoi{10.1051/0004-6361/201014250}

\bibitem[{{Bergemann} {et~al.}(2019){Bergemann}, {Gallagher}, {Eitner},
  {Bautista}, {Collet}, {Yakovleva}, {Mayriedl}, {Plez}, {Carlsson},
  {Leenaarts}, {Belyaev}, \& {Hansen}}]{bergemann2019}
{Bergemann}, M., {Gallagher}, A.~J., {Eitner}, P., {et~al.} 2019, \aap, 631,
  A80, \dodoi{10.1051/0004-6361/201935811}

\bibitem[{{Bernstein} {et~al.}(2003){Bernstein}, {Shectman}, {Gunnels},
  {Mochnacki}, \& {Athey}}]{bernstein2003}
{Bernstein}, R., {Shectman}, S.~A., {Gunnels}, S.~M., {Mochnacki}, S., \&
  {Athey}, A.~E. 2003, in Society of Photo-Optical Instrumentation Engineers
  (SPIE) Conference Series, Vol. 4841, \procspie, ed. M.~{Iye} \& A.~F.~M.
  {Moorwood}, 1694--1704, \dodoi{10.1117/12.461502}

\bibitem[{{Blanco-Cuaresma}(2019)}]{blanco-cuaresma2019}
{Blanco-Cuaresma}, S. 2019, \mnras, 486, 2075, \dodoi{10.1093/mnras/stz549}

\bibitem[{{Blanco-Cuaresma} {et~al.}(2014){Blanco-Cuaresma}, {Soubiran},
  {Heiter}, \& {Jofr{\'e}}}]{blanco-cuaresma2014}
{Blanco-Cuaresma}, S., {Soubiran}, C., {Heiter}, U., \& {Jofr{\'e}}, P. 2014,
  \aap, 569, A111, \dodoi{10.1051/0004-6361/201423945}

\bibitem[{{Bland-Hawthorn} \& {Gerhard}(2016)}]{blandhawthorn2016}
{Bland-Hawthorn}, J., \& {Gerhard}, O. 2016, \araa, 54, 529,
  \dodoi{10.1146/annurev-astro-081915-023441}

\bibitem[{{Bonifacio} {et~al.}(2009{\natexlab{a}}){Bonifacio}, {Spite},
  {Cayrel}, {Hill}, {Spite}, {Fran{\c c}ois}, {Plez}, {Ludwig}, {Caffau},
  {Molaro}, {Depagne}, {Andersen}, {Barbuy}, {Beers}, {Nordstr{\"o}m}, \&
  {Primas}}]{bonifacio2009}
{Bonifacio}, P., {Spite}, M., {Cayrel}, R., {et~al.} 2009{\natexlab{a}}, \aap,
  501, 519, \dodoi{10.1051/0004-6361/200810610}

\bibitem[{{Bonifacio} {et~al.}(2009{\natexlab{b}}){Bonifacio}, {Spite},
  {Cayrel}, {Hill}, {Spite}, {Fran{\c{c}}ois}, {Plez}, {Ludwig}, {Caffau},
  {Molaro}, {Depagne}, {Andersen}, {Barbuy}, {Beers}, {Nordstr{\"o}m}, \&
  {Primas}}]{bonifacio2009b}
---. 2009{\natexlab{b}}, \aap, 501, 519, \dodoi{10.1051/0004-6361/200810610}

\bibitem[{{Bovy}(2015)}]{bovy2015}
{Bovy}, J. 2015, \apjs, 216, 29, \dodoi{10.1088/0067-0049/216/2/29}

\bibitem[{{Brook} {et~al.}(2007){Brook}, {Kawata}, {Scannapieco}, {Martel}, \&
  {Gibson}}]{brook2007}
{Brook}, C.~B., {Kawata}, D., {Scannapieco}, E., {Martel}, H., \& {Gibson},
  B.~K. 2007, \apj, 661, 10, \dodoi{10.1086/511514}

\bibitem[{{Casagrande} {et~al.}(2010){Casagrande}, {Ram{\'\i}rez},
  {Mel{\'e}ndez}, {Bessell}, \& {Asplund}}]{casagrande2010}
{Casagrande}, L., {Ram{\'\i}rez}, I., {Mel{\'e}ndez}, J., {Bessell}, M., \&
  {Asplund}, M. 2010, \aap, 512, A54, \dodoi{10.1051/0004-6361/200913204}

\bibitem[{{Casey}(2016)}]{casey2016}
{Casey}, A.~R. 2016, \apjs, 223, 8, \dodoi{10.3847/0067-0049/223/1/8}

\bibitem[{{Casey} \& {Schlaufman}(2015)}]{casey2015}
{Casey}, A.~R., \& {Schlaufman}, K.~C. 2015, \apj, 809, 110,
  \dodoi{10.1088/0004-637X/809/2/110}

\bibitem[{{Casey} \& {Schlaufman}(2017)}]{casey2017}
---. 2017, \apj, 850, 179, \dodoi{10.3847/1538-4357/aa9079}

\bibitem[{{Castelli} \& {Kurucz}(2004)}]{castelli2004}
{Castelli}, F., \& {Kurucz}, R.~L. 2004, arXiv Astrophysics e-prints

\bibitem[{{Cayrel} {et~al.}(2004){Cayrel}, {Depagne}, {Spite}, {Hill}, {Spite},
  {Fran{\c{c}}ois}, {Plez}, {Beers}, {Primas}, {Andersen}, {Barbuy},
  {Bonifacio}, {Molaro}, \& {Nordstr{\"o}m}}]{cayrel2004}
{Cayrel}, R., {Depagne}, E., {Spite}, M., {et~al.} 2004, \aap, 416, 1117,
  \dodoi{10.1051/0004-6361:20034074}

\bibitem[{{Cescutti} {et~al.}(2006){Cescutti}, {Fran{\c{c}}ois}, {Matteucci},
  {Cayrel}, \& {Spite}}]{cescutti2006}
{Cescutti}, G., {Fran{\c{c}}ois}, P., {Matteucci}, F., {Cayrel}, R., \&
  {Spite}, M. 2006, \aap, 448, 557, \dodoi{10.1051/0004-6361:20053622}

\bibitem[{{Chieffi} \& {Limongi}(2004)}]{chieffi2004}
{Chieffi}, A., \& {Limongi}, M. 2004, \apj, 608, 405, \dodoi{10.1086/392523}

\bibitem[{{Churchwell} {et~al.}(2009){Churchwell}, {Babler}, {Meade},
  {Whitney}, {Benjamin}, {Indebetouw}, {Cyganowski}, {Robitaille}, {Povich},
  {Watson}, \& {Bracker}}]{churchwell2009}
{Churchwell}, E., {Babler}, B.~L., {Meade}, M.~R., {et~al.} 2009, \pasp, 121,
  213, \dodoi{10.1086/597811}

\bibitem[{{Clayton}(2003)}]{clayton2003}
{Clayton}, D. 2003, {Handbook of Isotopes in the Cosmos}

\bibitem[{{Cunha} \& {Smith}(2006)}]{cunha2006}
{Cunha}, K., \& {Smith}, V.~V. 2006, \apj, 651, 491, \dodoi{10.1086/507673}

\bibitem[{{de los Reyes} {et~al.}(2020){de los Reyes}, {Kirby}, {Seitenzahl},
  \& {Shen}}]{reyes2020}
{de los Reyes}, M. A.~C., {Kirby}, E.~N., {Seitenzahl}, I.~R., \& {Shen}, K.~J.
  2020, \apj, 891, 85, \dodoi{10.3847/1538-4357/ab736f}

\bibitem[{{Diemand} {et~al.}(2005){Diemand}, {Madau}, \& {Moore}}]{diemand2005}
{Diemand}, J., {Madau}, P., \& {Moore}, B. 2005, \mnras, 364, 367,
  \dodoi{10.1111/j.1365-2966.2005.09604.x}

\bibitem[{{Dotter} {et~al.}(2007){Dotter}, {Chaboyer}, {Jevremovi{\'c}},
  {Baron}, {Ferguson}, {Sarajedini}, \& {Anderson}}]{dotter2007}
{Dotter}, A., {Chaboyer}, B., {Jevremovi{\'c}}, D., {et~al.} 2007, \aj, 134,
  376, \dodoi{10.1086/517915}

\bibitem[{{Dotter} {et~al.}(2008){Dotter}, {Chaboyer}, {Jevremovi{\'c}},
  {Kostov}, {Baron}, \& {Ferguson}}]{dotter2008}
---. 2008, \apjs, 178, 89, \dodoi{10.1086/589654}

\bibitem[{{Eitner} {et~al.}(2020){Eitner}, {Bergemann}, {Hansen}, {Cescutti},
  {Seitenzahl}, {Larsen}, \& {Plez}}]{eitner2020}
{Eitner}, P., {Bergemann}, M., {Hansen}, C.~J., {et~al.} 2020, \aap, 635, A38,
  \dodoi{10.1051/0004-6361/201936603}

\bibitem[{{Feroz} \& {Hobson}(2008)}]{feroz2008}
{Feroz}, F., \& {Hobson}, M.~P. 2008, \mnras, 384, 449,
  \dodoi{10.1111/j.1365-2966.2007.12353.x}

\bibitem[{{Feroz} {et~al.}(2009){Feroz}, {Hobson}, \& {Bridges}}]{feroz2009}
{Feroz}, F., {Hobson}, M.~P., \& {Bridges}, M. 2009, \mnras, 398, 1601,
  \dodoi{10.1111/j.1365-2966.2009.14548.x}

\bibitem[{{Feroz} {et~al.}(2019){Feroz}, {Hobson}, {Cameron}, \&
  {Pettitt}}]{feroz2019}
{Feroz}, F., {Hobson}, M.~P., {Cameron}, E., \& {Pettitt}, A.~N. 2019, The Open
  Journal of Astrophysics, 2, 10, \dodoi{10.21105/astro.1306.2144}

\bibitem[{{Fink} {et~al.}(2014){Fink}, {Kromer}, {Seitenzahl},
  {Ciaraldi-Schoolmann}, {R{\"o}pke}, {Sim}, {Pakmor}, {Ruiter}, \&
  {Hillebrandt}}]{fink2014}
{Fink}, M., {Kromer}, M., {Seitenzahl}, I.~R., {et~al.} 2014, \mnras, 438,
  1762, \dodoi{10.1093/mnras/stt2315}

\bibitem[{{Frebel} {et~al.}(2013){Frebel}, {Casey}, {Jacobson}, \&
  {Yu}}]{frebel2013}
{Frebel}, A., {Casey}, A.~R., {Jacobson}, H.~R., \& {Yu}, Q. 2013, \apj, 769,
  57, \dodoi{10.1088/0004-637X/769/1/57}

\bibitem[{{Gaia Collaboration} {et~al.}(2016){Gaia Collaboration}, {Prusti},
  {de Bruijne}, {Brown}, {Vallenari}, {Babusiaux}, {Bailer-Jones}, {Bastian},
  {Biermann}, {Evans}, {Eyer}, {Jansen}, {Jordi}, {Klioner}, {Lammers},
  {Lindegren}, {Luri}, {Mignard}, {Milligan}, {Panem}, {Poinsignon},
  {Pourbaix}, {Randich}, {Sarri}, {Sartoretti}, {Siddiqui}, {Soubiran},
  {Valette}, {van Leeuwen}, {Walton}, {Aerts}, {Arenou}, {Cropper}, {Drimmel},
  {H{\o}g}, {Katz}, {Lattanzi}, {O'Mullane}, {Grebel}, {Holland}, {Huc},
  {Passot}, {Bramante}, {Cacciari}, {Casta{\~n}eda}, {Chaoul}, {Cheek}, {De
  Angeli}, {Fabricius}, {Guerra}, {Hern{\'a}ndez}, {Jean-Antoine-Piccolo},
  {Masana}, {Messineo}, {Mowlavi}, {Nienartowicz}, {Ord{\'o}{\~n}ez-Blanco},
  {Panuzzo}, {Portell}, {Richards}, {Riello}, {Seabroke}, {Tanga},
  {Th{\'e}venin}, {Torra}, {Els}, {Gracia-Abril}, {Comoretto},
  {Garcia-Reinaldos}, {Lock}, {Mercier}, {Altmann}, {Andrae}, {Astraatmadja},
  {Bellas-Velidis}, {Benson}, {Berthier}, {Blomme}, {Busso}, {Carry},
  {Cellino}, {Clementini}, {Cowell}, {Creevey}, {Cuypers}, {Davidson}, {De
  Ridder}, {de Torres}, {Delchambre}, {Dell'Oro}, {Ducourant}, {Fr{\'e}mat},
  {Garc{\'\i}a-Torres}, {Gosset}, {Halbwachs}, {Hambly}, {Harrison}, {Hauser},
  {Hestroffer}, {Hodgkin}, {Huckle}, {Hutton}, {Jasniewicz}, {Jordan},
  {Kontizas}, {Korn}, {Lanzafame}, {Manteiga}, {Moitinho}, {Muinonen},
  {Osinde}, {Pancino}, {Pauwels}, {Petit}, {Recio-Blanco}, {Robin}, {Sarro},
  {Siopis}, {Smith}, {Smith}, {Sozzetti}, {Thuillot}, {van Reeven}, {Viala},
  {Abbas}, {Abreu Aramburu}, {Accart}, {Aguado}, {Allan}, {Allasia},
  {Altavilla}, {{\'A}lvarez}, {Alves}, {Anderson}, {Andrei}, {Anglada Varela},
  {Antiche}, {Antoja}, {Ant{\'o}n}, {Arcay}, {Atzei}, {Ayache}, {Bach},
  {Baker}, {Balaguer-N{\'u}{\~n}ez}, {Barache}, {Barata}, {Barbier}, {Barblan},
  {Baroni}, {Barrado y Navascu{\'e}s}, {Barros}, {Barstow}, {Becciani},
  {Bellazzini}, {Bellei}, {Bello Garc{\'\i}a}, {Belokurov}, {Bendjoya},
  {Berihuete}, {Bianchi}, {Bienaym{\'e}}, {Billebaud}, {Blagorodnova},
  {Blanco-Cuaresma}, {Boch}, {Bombrun}, {Borrachero}, {Bouquillon}, {Bourda},
  {Bouy}, {Bragaglia}, {Breddels}, {Brouillet}, {Br{\"u}semeister},
  {Bucciarelli}, {Budnik}, {Burgess}, {Burgon}, {Burlacu}, {Busonero}, {Buzzi},
  {Caffau}, {Cambras}, {Campbell}, {Cancelliere}, {Cantat-Gaudin}, {Carlucci},
  {Carrasco}, {Castellani}, {Charlot}, {Charnas}, {Charvet}, {Chassat},
  {Chiavassa}, {Clotet}, {Cocozza}, {Collins}, {Collins}, {Costigan}, {Crifo},
  {Cross}, {Crosta}, {Crowley}, {Dafonte}, {Damerdji}, {Dapergolas}, {David},
  {David}, {De Cat}, {de Felice}, {de Laverny}, {De Luise}, {De March}, {de
  Martino}, {de Souza}, {Debosscher}, {del Pozo}, {Delbo}, {Delgado},
  {Delgado}, {di Marco}, {Di Matteo}, {Diakite}, {Distefano}, {Dolding}, {Dos
  Anjos}, {Drazinos}, {Dur{\'a}n}, {Dzigan}, {Ecale}, {Edvardsson}, {Enke},
  {Erdmann}, {Escolar}, {Espina}, {Evans}, {Eynard Bontemps}, {Fabre},
  {Fabrizio}, {Faigler}, {Falc{\~a}o}, {Farr{\`a}s Casas}, {Faye}, {Federici},
  {Fedorets}, {Fern{\'a}ndez-Hern{\'a}ndez}, {Fernique}, {Fienga}, {Figueras},
  {Filippi}, {Findeisen}, {Fonti}, {Fouesneau}, {Fraile}, {Fraser}, {Fuchs},
  {Furnell}, {Gai}, {Galleti}, {Galluccio}, {Garabato}, {Garc{\'\i}a-Sedano},
  {Gar{\'e}}, {Garofalo}, {Garralda}, {Gavras}, {Gerssen}, {Geyer}, {Gilmore},
  {Girona}, {Giuffrida}, {Gomes}, {Gonz{\'a}lez-Marcos},
  {Gonz{\'a}lez-N{\'u}{\~n}ez}, {Gonz{\'a}lez-Vidal}, {Granvik}, {Guerrier},
  {Guillout}, {Guiraud}, {G{\'u}rpide}, {Guti{\'e}rrez-S{\'a}nchez}, {Guy},
  {Haigron}, {Hatzidimitriou}, {Haywood}, {Heiter}, {Helmi}, {Hobbs},
  {Hofmann}, {Holl}, {Holland }, {Hunt}, {Hypki}, {Icardi}, {Irwin}, {Jevardat
  de Fombelle}, {Jofr{\'e}}, {Jonker}, {Jorissen}, {Julbe}, {Karampelas},
  {Kochoska}, {Kohley}, {Kolenberg}, {Kontizas}, {Koposov}, {Kordopatis},
  {Koubsky}, {Kowalczyk}, {Krone-Martins}, {Kudryashova}, {Kull}, {Bachchan},
  {Lacoste-Seris}, {Lanza}, {Lavigne}, {Le Poncin-Lafitte}, {Lebreton},
  {Lebzelter}, {Leccia}, {Leclerc}, {Lecoeur-Taibi}, {Lemaitre}, {Lenhardt},
  {Leroux}, {Liao}, {Licata}, {Lindstr{\o}m}, {Lister}, {Livanou}, {Lobel},
  {L{\"o}ffler}, {L{\'o}pez}, {Lopez-Lozano}, {Lorenz}, {Loureiro},
  {MacDonald}, {Magalh{\~a}es Fernandes}, {Managau}, {Mann}, {Mantelet},
  {Marchal}, {Marchant}, {Marconi}, {Marie}, {Marinoni}, {Marrese},
  {Marschalk{\'o}}, {Marshall}, {Mart{\'\i}n-Fleitas}, {Martino}, {Mary},
  {Matijevi{\v{c}}}, {Mazeh}, {McMillan}, {Messina}, {Mestre}, {Michalik},
  {Millar}, {Miranda}, {Molina}, {Molinaro}, {Molinaro}, {Moln{\'a}r},
  {Moniez}, {Montegriffo}, {Monteiro}, {Mor}, {Mora}, {Morbidelli}, {Morel},
  {Morgenthaler}, {Morley}, {Morris}, {Mulone}, {Muraveva}, {Musella},
  {Narbonne}, {Nelemans}, {Nicastro}, {Noval}, {Ord{\'e}novic},
  {Ordieres-Mer{\'e}}, {Osborne}, {Pagani}, {Pagano}, {Pailler}, {Palacin},
  {Palaversa}, {Parsons}, {Paulsen}, {Pecoraro}, {Pedrosa}, {Pentik{\"a}inen},
  {Pereira}, {Pichon}, {Piersimoni}, {Pineau}, {Plachy}, {Plum}, {Poujoulet},
  {Pr{\v{s}}a}, {Pulone}, {Ragaini}, {Rago}, {Rambaux}, {Ramos-Lerate},
  {Ranalli}, {Rauw}, {Read}, {Regibo}, {Renk}, {Reyl{\'e}}, {Ribeiro},
  {Rimoldini}, {Ripepi}, {Riva}, {Rixon}, {Roelens}, {Romero-G{\'o}mez},
  {Rowell}, {Royer}, {Rudolph}, {Ruiz-Dern}, {Sadowski}, {Sagrist{\`a}
  Sell{\'e}s}, {Sahlmann}, {Salgado}, {Salguero}, {Sarasso}, {Savietto},
  {Schnorhk}, {Schultheis}, {Sciacca}, {Segol}, {Segovia}, {Segransan},
  {Serpell}, {Shih}, {Smareglia}, {Smart}, {Smith}, {Solano}, {Solitro},
  {Sordo}, {Soria Nieto}, {Souchay}, {Spagna}, {Spoto}, {Stampa}, {Steele},
  {Steidelm{\"u}ller}, {Stephenson}, {Stoev}, {Suess}, {S{\"u}veges}, {Surdej},
  {Szabados}, {Szegedi-Elek}, {Tapiador}, {Taris}, {Tauran}, {Taylor},
  {Teixeira}, {Terrett}, {Tingley}, {Trager}, {Turon}, {Ulla}, {Utrilla},
  {Valentini}, {van Elteren}, {Van Hemelryck}, {van Leeuwen}, {Varadi},
  {Vecchiato}, {Veljanoski}, {Via}, {Vicente}, {Vogt}, {Voss}, {Votruba},
  {Voutsinas}, {Walmsley}, {Weiler}, {Weingrill}, {Werner}, {Wevers},
  {Whitehead}, {Wyrzykowski}, {Yoldas}, {{\v{Z}}erjal}, {Zucker}, {Zurbach},
  {Zwitter}, {Alecu}, {Allen}, {Allende Prieto}, {Amorim},
  {Anglada-Escud{\'e}}, {Arsenijevic}, {Azaz}, {Balm}, {Beck}, {Bernstein},
  {Bigot}, {Bijaoui}, {Blasco}, {Bonfigli}, {Bono}, {Boudreault}, {Bressan},
  {Brown}, {Brunet}, {Bunclark}, {Buonanno}, {Butkevich}, {Carret}, {Carrion},
  {Chemin}, {Ch{\'e}reau}, {Corcione}, {Darmigny}, {de Boer}, {de Teodoro}, {de
  Zeeuw}, {Delle Luche}, {Domingues}, {Dubath}, {Fodor}, {Fr{\'e}zouls},
  {Fries}, {Fustes}, {Fyfe}, {Gallardo}, {Gallegos}, {Gardiol}, {Gebran},
  {Gomboc}, {G{\'o}mez}, {Grux}, {Gueguen}, {Heyrovsky}, {Hoar}, {Iannicola},
  {Isasi Parache}, {Janotto}, {Joliet}, {Jonckheere}, {Keil}, {Kim},
  {Klagyivik}, {Klar}, {Knude}, {Kochukhov}, {Kolka}, {Kos}, {Kutka}, {Lainey},
  {LeBouquin}, {Liu}, {Loreggia}, {Makarov}, {Marseille}, {Martayan},
  {Martinez-Rubi}, {Massart}, {Meynadier}, {Mignot}, {Munari}, {Nguyen},
  {Nordlander}, {Ocvirk}, {O'Flaherty}, {Olias Sanz}, {Ortiz}, {Osorio},
  {Oszkiewicz}, {Ouzounis}, {Palmer}, {Park}, {Pasquato}, {Peltzer}, {Peralta},
  {P{\'e}turaud}, {Pieniluoma}, {Pigozzi}, {Poels}, {Prat}, {Prod'homme},
  {Raison}, {Rebordao}, {Risquez}, {Rocca-Volmerange}, {Rosen}, {Ruiz-Fuertes},
  {Russo}, {Sembay}, {Serraller Vizcaino}, {Short}, {Siebert}, {Silva},
  {Sinachopoulos}, {Slezak}, {Soffel}, {Sosnowska}, {Strai{\v{z}}ys}, {ter
  Linden}, {Terrell}, {Theil}, {Tiede}, {Troisi}, {Tsalmantza}, {Tur},
  {Vaccari}, {Vachier}, {Valles}, {Van Hamme}, {Veltz}, {Virtanen}, {Wallut},
  {Wichmann}, {Wilkinson}, {Ziaeepour}, \& {Zschocke}}]{gaia2016}
{Gaia Collaboration}, {Prusti}, T., {de Bruijne}, J.~H.~J., {et~al.} 2016,
  \aap, 595, A1, \dodoi{10.1051/0004-6361/201629272}

\bibitem[{{Gaia Collaboration} {et~al.}(2018){Gaia Collaboration}, {Brown},
  {Vallenari}, {Prusti}, {de Bruijne}, {Babusiaux}, {Bailer-Jones}, {Biermann},
  {Evans}, {Eyer}, {Jansen}, {Jordi}, {Klioner}, {Lammers}, {Lindegren},
  {Luri}, {Mignard}, {Panem}, {Pourbaix}, {Randich}, {Sartoretti}, {Siddiqui},
  {Soubiran}, {van Leeuwen}, {Walton}, {Arenou}, {Bastian}, {Cropper},
  {Drimmel}, {Katz}, {Lattanzi}, {Bakker}, {Cacciari}, {Casta{\~n}eda},
  {Chaoul}, {Cheek}, {De Angeli}, {Fabricius}, {Guerra}, {Holl}, {Masana},
  {Messineo}, {Mowlavi}, {Nienartowicz}, {Panuzzo}, {Portell}, {Riello},
  {Seabroke}, {Tanga}, {Th{\'e}venin}, {Gracia-Abril}, {Comoretto},
  {Garcia-Reinaldos}, {Teyssier}, {Altmann}, {Andrae}, {Audard},
  {Bellas-Velidis}, {Benson}, {Berthier}, {Blomme}, {Burgess}, {Busso},
  {Carry}, {Cellino}, {Clementini}, {Clotet}, {Creevey}, {Davidson}, {De
  Ridder}, {Delchambre}, {Dell'Oro}, {Ducourant},
  {Fern{\'a}ndez-Hern{\'a}ndez}, {Fouesneau}, {Fr{\'e}mat}, {Galluccio},
  {Garc{\'\i}a-Torres}, {Gonz{\'a}lez-N{\'u}{\~n}ez}, {Gonz{\'a}lez-Vidal},
  {Gosset}, {Guy}, {Halbwachs}, {Hambly}, {Harrison}, {Hern{\'a}ndez},
  {Hestroffer}, {Hodgkin}, {Hutton}, {Jasniewicz}, {Jean-Antoine-Piccolo},
  {Jordan}, {Korn}, {Krone-Martins}, {Lanzafame}, {Lebzelter}, {L{\"o}ffler},
  {Manteiga}, {Marrese}, {Mart{\'\i}n-Fleitas}, {Moitinho}, {Mora}, {Muinonen},
  {Osinde}, {Pancino}, {Pauwels}, {Petit}, {Recio-Blanco}, {Richards},
  {Rimoldini}, {Robin}, {Sarro}, {Siopis}, {Smith}, {Sozzetti}, {S{\"u}veges},
  {Torra}, {van Reeven}, {Abbas}, {Abreu Aramburu}, {Accart}, {Aerts},
  {Altavilla}, {{\'A}lvarez}, {Alvarez}, {Alves}, {Anderson}, {Andrei},
  {Anglada Varela}, {Antiche}, {Antoja}, {Arcay}, {Astraatmadja}, {Bach},
  {Baker}, {Balaguer-N{\'u}{\~n}ez}, {Balm}, {Barache}, {Barata}, {Barbato},
  {Barblan}, {Barklem}, {Barrado}, {Barros}, {Barstow}, {Bartholom{\'e}
  Mu{\~n}oz}, {Bassilana}, {Becciani}, {Bellazzini}, {Berihuete}, {Bertone},
  {Bianchi}, {Bienaym{\'e}}, {Blanco-Cuaresma}, {Boch}, {Boeche}, {Bombrun},
  {Borrachero}, {Bossini}, {Bouquillon}, {Bourda}, {Bragaglia}, {Bramante},
  {Breddels}, {Bressan}, {Brouillet}, {Br{\"u}semeister}, {Brugaletta},
  {Bucciarelli}, {Burlacu}, {Busonero}, {Butkevich}, {Buzzi}, {Caffau},
  {Cancelliere}, {Cannizzaro}, {Cantat-Gaudin}, {Carballo}, {Carlucci},
  {Carrasco}, {Casamiquela}, {Castellani}, {Castro-Ginard}, {Charlot},
  {Chemin}, {Chiavassa}, {Cocozza}, {Costigan}, {Cowell}, {Crifo}, {Crosta},
  {Crowley}, {Cuypers}, {Dafonte}, {Damerdji}, {Dapergolas}, {David}, {David},
  {de Laverny}, {De Luise}, {De March}, {de Martino}, {de Souza}, {de Torres},
  {Debosscher}, {del Pozo}, {Delbo}, {Delgado}, {Delgado}, {Di Matteo},
  {Diakite}, {Diener}, {Distefano}, {Dolding}, {Drazinos}, {Dur{\'a}n},
  {Edvardsson}, {Enke}, {Eriksson}, {Esquej}, {Eynard Bontemps}, {Fabre},
  {Fabrizio}, {Faigler}, {Falc{\~a}o}, {Farr{\`a}s Casas}, {Federici},
  {Fedorets}, {Fernique}, {Figueras}, {Filippi}, {Findeisen}, {Fonti},
  {Fraile}, {Fraser}, {Fr{\'e}zouls}, {Gai}, {Galleti}, {Garabato},
  {Garc{\'\i}a-Sedano}, {Garofalo}, {Garralda}, {Gavel}, {Gavras}, {Gerssen},
  {Geyer}, {Giacobbe}, {Gilmore}, {Girona}, {Giuffrida}, {Glass}, {Gomes},
  {Granvik}, {Gueguen}, {Guerrier}, {Guiraud}, {Guti{\'e}rrez-S{\'a}nchez},
  {Haigron}, {Hatzidimitriou}, {Hauser}, {Haywood}, {Heiter}, {Helmi}, {Heu},
  {Hilger}, {Hobbs}, {Hofmann}, {Holland}, {Huckle}, {Hypki}, {Icardi},
  {Jan{\ss}en}, {Jevardat de Fombelle}, {Jonker}, {Juh{\'a}sz}, {Julbe},
  {Karampelas}, {Kewley}, {Klar}, {Kochoska}, {Kohley}, {Kolenberg},
  {Kontizas}, {Kontizas}, {Koposov}, {Kordopatis}, {Kostrzewa-Rutkowska},
  {Koubsky}, {Lambert}, {Lanza}, {Lasne}, {Lavigne}, {Le Fustec}, {Le
  Poncin-Lafitte}, {Lebreton}, {Leccia}, {Leclerc}, {Lecoeur-Taibi},
  {Lenhardt}, {Leroux}, {Liao}, {Licata}, {Lindstr{\o}m}, {Lister}, {Livanou},
  {Lobel}, {L{\'o}pez}, {Managau}, {Mann}, {Mantelet}, {Marchal}, {Marchant},
  {Marconi}, {Marinoni}, {Marschalk{\'o}}, {Marshall}, {Martino}, {Marton},
  {Mary}, {Massari}, {Matijevi{\v{c}}}, {Mazeh}, {McMillan}, {Messina},
  {Michalik}, {Millar}, {Molina}, {Molinaro}, {Moln{\'a}r}, {Montegriffo},
  {Mor}, {Morbidelli}, {Morel}, {Morris}, {Mulone}, {Muraveva}, {Musella},
  {Nelemans}, {Nicastro}, {Noval}, {O'Mullane}, {Ord{\'e}novic},
  {Ord{\'o}{\~n}ez-Blanco}, {Osborne}, {Pagani}, {Pagano}, {Pailler},
  {Palacin}, {Palaversa}, {Panahi}, {Pawlak}, {Piersimoni}, {Pineau}, {Plachy},
  {Plum}, {Poggio}, {Poujoulet}, {Pr{\v{s}}a}, {Pulone}, {Racero}, {Ragaini},
  {Rambaux}, {Ramos-Lerate}, {Regibo}, {Reyl{\'e}}, {Riclet}, {Ripepi}, {Riva},
  {Rivard}, {Rixon}, {Roegiers}, {Roelens}, {Romero-G{\'o}mez}, {Rowell},
  {Royer}, {Ruiz-Dern}, {Sadowski}, {Sagrist{\`a} Sell{\'e}s}, {Sahlmann},
  {Salgado}, {Salguero}, {Sanna}, {Santana-Ros}, {Sarasso}, {Savietto},
  {Schultheis}, {Sciacca}, {Segol}, {Segovia}, {S{\'e}gransan}, {Shih},
  {Siltala}, {Silva}, {Smart}, {Smith}, {Solano}, {Solitro}, {Sordo}, {Soria
  Nieto}, {Souchay}, {Spagna}, {Spoto}, {Stampa}, {Steele},
  {Steidelm{\"u}ller}, {Stephenson}, {Stoev}, {Suess}, {Surdej}, {Szabados},
  {Szegedi-Elek}, {Tapiador}, {Taris}, {Tauran}, {Taylor}, {Teixeira},
  {Terrett}, {Teyssand ier}, {Thuillot}, {Titarenko}, {Torra Clotet}, {Turon},
  {Ulla}, {Utrilla}, {Uzzi}, {Vaillant}, {Valentini}, {Valette}, {van Elteren},
  {Van Hemelryck}, {van Leeuwen}, {Vaschetto}, {Vecchiato}, {Veljanoski},
  {Viala}, {Vicente}, {Vogt}, {von Essen}, {Voss}, {Votruba}, {Voutsinas},
  {Walmsley}, {Weiler}, {Wertz}, {Wevers}, {Wyrzykowski}, {Yoldas},
  {{\v{Z}}erjal}, {Ziaeepour}, {Zorec}, {Zschocke}, {Zucker}, {Zurbach}, \&
  {Zwitter}}]{gaia2018}
{Gaia Collaboration}, {Brown}, A.~G.~A., {Vallenari}, A., {et~al.} 2018, \aap,
  616, A1, \dodoi{10.1051/0004-6361/201833051}

\bibitem[{{Gao} {et~al.}(2010){Gao}, {Theuns}, {Frenk}, {Jenkins}, {Helly},
  {Navarro}, {Springel}, \& {White}}]{gao2010}
{Gao}, L., {Theuns}, T., {Frenk}, C.~S., {et~al.} 2010, \mnras, 403, 1283,
  \dodoi{10.1111/j.1365-2966.2009.16225.x}

\bibitem[{{Garc{\'\i}a P{\'e}rez} {et~al.}(2013){Garc{\'\i}a P{\'e}rez},
  {Cunha}, {Shetrone}, {Majewski}, {Johnson}, {Smith}, {Schiavon}, {Holtzman},
  {Nidever}, {Zasowski}, {Allende Prieto}, {Beers}, {Bizyaev}, {Ebelke},
  {Eisenstein}, {Frinchaboy}, {Girardi}, {Hearty}, {Malanushenko},
  {Malanushenko}, {Meszaros}, {O'Connell}, {Oravetz}, {Pan}, {Robin},
  {Schneider}, {Schultheis}, {Skrutskie}, {Simmonsand }, \&
  {Wilson}}]{garcia-perez2013}
{Garc{\'\i}a P{\'e}rez}, A.~E., {Cunha}, K., {Shetrone}, M., {et~al.} 2013,
  \apjl, 767, L9, \dodoi{10.1088/2041-8205/767/1/L9}

\bibitem[{{Gargiulo} {et~al.}(2017){Gargiulo}, {Cora}, {Vega-Mart{\'\i}nez},
  {Gonzalez}, {Zoccali}, {Gonz{\'a}lez}, {Ruiz}, \& {Padilla}}]{gargiulo2017}
{Gargiulo}, I.~D., {Cora}, S.~A., {Vega-Mart{\'\i}nez}, C.~A., {et~al.} 2017,
  \mnras, 472, 4133, \dodoi{10.1093/mnras/stx2188}

\bibitem[{{Gonzalez} {et~al.}(2011){Gonzalez}, {Rejkuba}, {Zoccali}, {Valenti},
  \& {Minniti}}]{gonzalez2011}
{Gonzalez}, O.~A., {Rejkuba}, M., {Zoccali}, M., {Valenti}, E., \& {Minniti},
  D. 2011, \aap, 534, A3, \dodoi{10.1051/0004-6361/201117601}

\bibitem[{{Gonzalez} {et~al.}(2012){Gonzalez}, {Rejkuba}, {Zoccali}, {Valenti},
  {Minniti}, {Schultheis}, {Tobar}, \& {Chen}}]{gonzalez2012}
{Gonzalez}, O.~A., {Rejkuba}, M., {Zoccali}, M., {et~al.} 2012, \aap, 543, A13,
  \dodoi{10.1051/0004-6361/201219222}

\bibitem[{{Gravity Collaboration} {et~al.}(2018){Gravity Collaboration},
  {Abuter}, {Amorim}, {Anugu}, {Baub{\"o}ck}, {Benisty}, {Berger}, {Blind},
  {Bonnet}, {Brandner}, {Buron}, {Collin}, {Chapron}, {Cl{\'e}net}, {Coud{\'e}
  Du Foresto}, {de Zeeuw}, {Deen}, {Delplancke-Str{\"o}bele}, {Dembet},
  {Dexter}, {Duvert}, {Eckart}, {Eisenhauer}, {Finger}, {F{\"o}rster
  Schreiber}, {F{\'e}dou}, {Garcia}, {Garcia Lopez}, {Gao}, {Gendron},
  {Genzel}, {Gillessen}, {Gordo}, {Habibi}, {Haubois}, {Haug}, {Hau{\ss}mann},
  {Henning}, {Hippler}, {Horrobin}, {Hubert}, {Hubin}, {Jimenez Rosales},
  {Jochum}, {Jocou}, {Kaufer}, {Kellner}, {Kendrew}, {Kervella}, {Kok},
  {Kulas}, {Lacour}, {Lapeyr{\`e}re}, {Lazareff}, {Le Bouquin}, {L{\'e}na},
  {Lippa}, {Lenzen}, {M{\'e}rand}, {M{\"u}ler}, {Neumann}, {Ott}, {Palanca},
  {Paumard}, {Pasquini}, {Perraut}, {Perrin}, {Pfuhl}, {Plewa}, {Rabien},
  {Ram{\'\i}rez}, {Ramos}, {Rau}, {Rodr{\'\i}guez-Coira}, {Rohloff}, {Rousset},
  {Sanchez-Bermudez}, {Scheithauer}, {Sch{\"o}ller}, {Schuler}, {Spyromilio},
  {Straub}, {Straubmeier}, {Sturm}, {Tacconi}, {Tristram}, {Vincent}, {von
  Fellenberg}, {Wank}, {Waisberg}, {Widmann}, {Wieprecht}, {Wiest},
  {Wiezorrek}, {Woillez}, {Yazici}, {Ziegler}, \& {Zins}}]{gravity2018}
{Gravity Collaboration}, {Abuter}, R., {Amorim}, A., {et~al.} 2018, \aap, 615,
  L15, \dodoi{10.1051/0004-6361/201833718}

\bibitem[{{Griffen} {et~al.}(2018){Griffen}, {Dooley}, {Ji}, {O'Shea},
  {G{\'o}mez}, \& {Frebel}}]{griffen2018}
{Griffen}, B.~F., {Dooley}, G.~A., {Ji}, A.~P., {et~al.} 2018, \mnras, 474,
  443, \dodoi{10.1093/mnras/stx2749}

\bibitem[{{Grimmett} {et~al.}(2019){Grimmett}, {Karakas}, {Heger}, \&
  {M{\"u}ller}}]{grimmett2019}
{Grimmett}, J.~J., {Karakas}, A.~I., {Heger}, A., \& {M{\"u}ller}, B. 2019,
  arXiv e-prints, arXiv:1911.05901.
\newblock \doarXiv{1911.05901}

\bibitem[{{Gustafsson} {et~al.}(2008){Gustafsson}, {Edvardsson}, {Eriksson},
  {J{\o}rgensen}, {Nordlund}, \& {Plez}}]{gustafsson2008}
{Gustafsson}, B., {Edvardsson}, B., {Eriksson}, K., {et~al.} 2008, \aap, 486,
  951, \dodoi{10.1051/0004-6361:200809724}

\bibitem[{{Hambly} {et~al.}(2018){Hambly}, {Cropper}, {Boudreault}, {Crowley},
  {Kohley}, {de Bruijne}, {Dolding}, {Fabricius}, {Seabroke}, {Davidson},
  {Rowell}, {Collins}, {Cross}, {Mart{\'\i}n-Fleitas}, {Baker}, {Smith},
  {Sartoretti}, {Marchal}, {Katz}, {De Angeli}, {Busso}, {Riello}, {Allende
  Prieto}, {Els}, {Corcione}, {Masana}, {Luri}, {Chassat}, {Fusero},
  {Pasquier}, {V{\'e}tel}, {Sarri}, \& {Gare}}]{hambly2018}
{Hambly}, N.~C., {Cropper}, M., {Boudreault}, S., {et~al.} 2018, \aap, 616,
  A15, \dodoi{10.1051/0004-6361/201832716}

\bibitem[{{Hansen} {et~al.}(2013){Hansen}, {Bergemann}, {Cescutti},
  {Fran{\c{c}}ois}, {Arcones}, {Karakas}, {Lind}, \& {Chiappini}}]{hansen2013}
{Hansen}, C.~J., {Bergemann}, M., {Cescutti}, G., {et~al.} 2013, \aap, 551,
  A57, \dodoi{10.1051/0004-6361/201220584}

\bibitem[{{Heger} \& {Woosley}(2010)}]{heger2010}
{Heger}, A., \& {Woosley}, S.~E. 2010, \apj, 724, 341,
  \dodoi{10.1088/0004-637X/724/1/341}

\bibitem[{{Howes} {et~al.}(2015){Howes}, {Casey}, {Asplund}, {Keller}, {Yong},
  {Nataf}, {Poleski}, {Lind}, {Kobayashi}, {Owen}, {Ness}, {Bessell}, {da
  Costa}, {Schmidt}, {Tisserand}, {Udalski}, {Szyma{\'n}ski}, {Soszy{\'n}ski},
  {Pietrzy{\'n}ski}, {Ulaczyk}, {Wyrzykowski}, {Pietrukowicz}, {Skowron},
  {Koz{\l}owski}, \& {Mr{\'o}z}}]{howes2015}
{Howes}, L.~M., {Casey}, A.~R., {Asplund}, M., {et~al.} 2015, \nat, 527, 484,
  \dodoi{10.1038/nature15747}

\bibitem[{{Howes} {et~al.}(2016){Howes}, {Asplund}, {Keller}, {Casey}, {Yong},
  {Lind}, {Frebel}, {Hays}, {Alves-Brito}, {Bessell}, {Casagrande}, {Marino},
  {Nataf}, {Owen}, {Da Costa}, {Schmidt}, \& {Tisserand}}]{howes2016}
{Howes}, L.~M., {Asplund}, M., {Keller}, S.~C., {et~al.} 2016, \mnras, 460,
  884, \dodoi{10.1093/mnras/stw1004}

\bibitem[{{Ishiyama} {et~al.}(2016){Ishiyama}, {Sudo}, {Yokoi}, {Hasegawa},
  {Tominaga}, \& {Susa}}]{ishiyama2016}
{Ishiyama}, T., {Sudo}, K., {Yokoi}, S., {et~al.} 2016, \apj, 826, 9,
  \dodoi{10.3847/0004-637X/826/1/9}

\bibitem[{{Jacobson} \& {Friel}(2013)}]{jacobson2013}
{Jacobson}, H.~R., \& {Friel}, E.~D. 2013, \aj, 145, 107,
  \dodoi{10.1088/0004-6256/145/4/107}

\bibitem[{{Ji} {et~al.}(2016){Ji}, {Frebel}, {Chiti}, \& {Simon}}]{ji2016a}
{Ji}, A.~P., {Frebel}, A., {Chiti}, A., \& {Simon}, J.~D. 2016, \nat, 531, 610,
  \dodoi{10.1038/nature17425}

\bibitem[{{Johnson} {et~al.}(2012){Johnson}, {Rich}, {Kobayashi}, \&
  {Fulbright}}]{johnson2012}
{Johnson}, C.~I., {Rich}, R.~M., {Kobayashi}, C., \& {Fulbright}, J.~P. 2012,
  \apj, 749, 175, \dodoi{10.1088/0004-637X/749/2/175}

\bibitem[{{Juri{\'c}} {et~al.}(2008){Juri{\'c}}, {Ivezi{\'c}}, {Brooks},
  {Lupton}, {Schlegel}, {Finkbeiner}, {Padmanabhan}, {Bond}, {Sesar},
  {Rockosi}, {Knapp}, {Gunn}, {Sumi}, {Schneider}, {Barentine}, {Brewington},
  {Brinkmann}, {Fukugita}, {Harvanek}, {Kleinman}, {Krzesinski}, {Long},
  {Neilsen}, {Nitta}, {Snedden}, \& {York}}]{juric2008}
{Juri{\'c}}, M., {Ivezi{\'c}}, {\v{Z}}., {Brooks}, A., {et~al.} 2008, \apj,
  673, 864, \dodoi{10.1086/523619}

\bibitem[{{Kelson}(2003)}]{kelson2003}
{Kelson}, D.~D. 2003, \pasp, 115, 688, \dodoi{10.1086/375502}

\bibitem[{{Kelson} {et~al.}(2000){Kelson}, {Illingworth}, {van Dokkum}, \&
  {Franx}}]{kelson2000}
{Kelson}, D.~D., {Illingworth}, G.~D., {van Dokkum}, P.~G., \& {Franx}, M.
  2000, \apj, 531, 159, \dodoi{10.1086/308445}

\bibitem[{{Kelson} {et~al.}(2014){Kelson}, {Williams}, {Dressler}, {McCarthy},
  {Shectman}, {Mulchaey}, {Villanueva}, {Crane}, \& {Quadri}}]{kelson2014}
{Kelson}, D.~D., {Williams}, R.~J., {Dressler}, A., {et~al.} 2014, \apj, 783,
  110, \dodoi{10.1088/0004-637X/783/2/110}

\bibitem[{{Kobayashi} {et~al.}(2006){Kobayashi}, {Umeda}, {Nomoto}, {Tominaga},
  \& {Ohkubo}}]{kobayashi2006}
{Kobayashi}, C., {Umeda}, H., {Nomoto}, K., {Tominaga}, N., \& {Ohkubo}, T.
  2006, \apj, 653, 1145, \dodoi{10.1086/508914}

\bibitem[{{Korn} {et~al.}(2003){Korn}, {Shi}, \& {Gehren}}]{korn2003}
{Korn}, A.~J., {Shi}, J., \& {Gehren}, T. 2003, \aap, 407, 691,
  \dodoi{10.1051/0004-6361:20030907}

\bibitem[{{Lamb} {et~al.}(2017){Lamb}, {Venn}, {Andersen}, {Oya}, {Shetrone},
  {Fattahi}, {Howes}, {Asplund}, {Lardi{\`e}re}, {Akiyama}, {Ono}, {Terada},
  {Hayano}, {Suzuki}, {Blain}, {Jackson}, {Correia}, {Youakim}, \&
  {Bradley}}]{lamb2017}
{Lamb}, M., {Venn}, K., {Andersen}, D., {et~al.} 2017, \mnras, 465, 3536,
  \dodoi{10.1093/mnras/stw2865}

\bibitem[{{Lind} {et~al.}(2011){Lind}, {Asplund}, {Barklem}, \&
  {Belyaev}}]{lind2011}
{Lind}, K., {Asplund}, M., {Barklem}, P.~S., \& {Belyaev}, A.~K. 2011, \aap,
  528, A103, \dodoi{10.1051/0004-6361/201016095}

\bibitem[{{Lindegren} {et~al.}(2018){Lindegren}, {Hern{\'a}ndez}, {Bombrun},
  {Klioner}, {Bastian}, {Ramos-Lerate}, {de Torres}, {Steidelm{\"u}ller},
  {Stephenson}, {Hobbs}, {Lammers}, {Biermann}, {Geyer}, {Hilger}, {Michalik},
  {Stampa}, {McMillan}, {Casta{\~n}eda}, {Clotet}, {Comoretto}, {Davidson},
  {Fabricius}, {Gracia}, {Hambly}, {Hutton}, {Mora}, {Portell}, {van Leeuwen},
  {Abbas}, {Abreu}, {Altmann}, {Andrei}, {Anglada}, {Balaguer-N{\'u}{\~n}ez},
  {Barache}, {Becciani}, {Bertone}, {Bianchi}, {Bouquillon}, {Bourda},
  {Br{\"u}semeister}, {Bucciarelli}, {Busonero}, {Buzzi}, {Cancelliere},
  {Carlucci}, {Charlot}, {Cheek}, {Crosta}, {Crowley}, {de Bruijne}, {de
  Felice}, {Drimmel}, {Esquej}, {Fienga}, {Fraile}, {Gai}, {Garralda},
  {Gonz{\'a}lez-Vidal}, {Guerra}, {Hauser}, {Hofmann}, {Holl}, {Jordan},
  {Lattanzi}, {Lenhardt}, {Liao}, {Licata}, {Lister}, {L{\"o}ffler},
  {Marchant}, {Martin-Fleitas}, {Messineo}, {Mignard}, {Morbidelli}, {Poggio},
  {Riva}, {Rowell}, {Salguero}, {Sarasso}, {Sciacca}, {Siddiqui}, {Smart},
  {Spagna}, {Steele}, {Taris}, {Torra}, {van Elteren}, {van Reeven}, \&
  {Vecchiato}}]{lindegren2018}
{Lindegren}, L., {Hern{\'a}ndez}, J., {Bombrun}, A., {et~al.} 2018, \aap, 616,
  A2, \dodoi{10.1051/0004-6361/201832727}

\bibitem[{{Livne} \& {Arnett}(1995)}]{livne1995}
{Livne}, E., \& {Arnett}, D. 1995, \apj, 452, 62, \dodoi{10.1086/176279}

\bibitem[{{Lucey} {et~al.}(2019){Lucey}, {Hawkins}, {Ness}, {Asplund},
  {Bensby}, {Casagrande}, {Feltzing}, {Freeman}, {Kobayashi}, \&
  {Marino}}]{lucey2019}
{Lucey}, M., {Hawkins}, K., {Ness}, M., {et~al.} 2019, \mnras, 488, 2283,
  \dodoi{10.1093/mnras/stz1847}

\bibitem[{{Luri} {et~al.}(2018){Luri}, {Brown}, {Sarro}, {Arenou},
  {Bailer-Jones}, {Castro-Ginard}, {de Bruijne}, {Prusti}, {Babusiaux}, \&
  {Delgado}}]{luri2018}
{Luri}, X., {Brown}, A.~G.~A., {Sarro}, L.~M., {et~al.} 2018, \aap, 616, A9,
  \dodoi{10.1051/0004-6361/201832964}

\bibitem[{{Mainzer} {et~al.}(2011){Mainzer}, {Grav}, {Bauer}, {Masiero},
  {McMillan}, {Cutri}, {Walker}, {Wright}, {Eisenhardt}, {Tholen}, {Spahr},
  {Jedicke}, {Denneau}, {DeBaun}, {Elsbury}, {Gautier}, {Gomillion}, {Hand},
  {Mo}, {Watkins}, {Wilkins}, {Bryngelson}, {Del Pino Molina}, {Desai},
  {G{\'o}mez Camus}, {Hidalgo}, {Konstantopoulos}, {Larsen}, {Maleszewski},
  {Malkan}, {Mauduit}, {Mullan}, {Olszewski}, {Pforr}, {Saro}, {Scotti}, \&
  {Wasserman}}]{mainzer2011}
{Mainzer}, A., {Grav}, T., {Bauer}, J., {et~al.} 2011, \apj, 743, 156,
  \dodoi{10.1088/0004-637X/743/2/156}

\bibitem[{{Mashonkina} \& {Belyaev}(2019)}]{mashonkina2019}
{Mashonkina}, L.~I., \& {Belyaev}, A.~K. 2019, Astronomy Letters, 45, 341,
  \dodoi{10.1134/S1063773719060033}

\bibitem[{{Meng} {et~al.}(2008){Meng}, {Chen}, \& {Han}}]{meng2008}
{Meng}, X., {Chen}, X., \& {Han}, Z. 2008, \aap, 487, 625,
  \dodoi{10.1051/0004-6361:20078841}

\bibitem[{{Miyamoto} \& {Nagai}(1975)}]{miyamoto1975}
{Miyamoto}, M., \& {Nagai}, R. 1975, \pasj, 27, 533

\bibitem[{{Morton}(2015)}]{morton2015}
{Morton}, T.~D. 2015, {isochrones: Stellar model grid package}.
\newblock \doeprint{1503.010}

\bibitem[{{Mucciarelli} \& {Bonifacio}(2020)}]{mucciarelli2020}
{Mucciarelli}, A., \& {Bonifacio}, P. 2020, arXiv e-prints, arXiv:2003.07390.
\newblock \doarXiv{2003.07390}

\bibitem[{{Navarro} {et~al.}(1996){Navarro}, {Frenk}, \& {White}}]{navarro1996}
{Navarro}, J.~F., {Frenk}, C.~S., \& {White}, S. D.~M. 1996, \apj, 462, 563,
  \dodoi{10.1086/177173}

\bibitem[{{Ness} {et~al.}(2013){Ness}, {Freeman}, {Athanassoula},
  {Wylie-de-Boer}, {Bland-Hawthorn}, {Asplund}, {Lewis}, {Yong}, {Lane}, \&
  {Kiss}}]{ness2013}
{Ness}, M., {Freeman}, K., {Athanassoula}, E., {et~al.} 2013, \mnras, 430, 836,
  \dodoi{10.1093/mnras/sts629}

\bibitem[{{Nishiyama} {et~al.}(2009){Nishiyama}, {Tamura}, {Hatano}, {Kato},
  {Tanab{\'e}}, {Sugitani}, \& {Nagata}}]{nishiyama2009}
{Nishiyama}, S., {Tamura}, M., {Hatano}, H., {et~al.} 2009, \apj, 696, 1407,
  \dodoi{10.1088/0004-637X/696/2/1407}

\bibitem[{{Ochsenbein} {et~al.}(2000){Ochsenbein}, {Bauer}, \&
  {Marcout}}]{ochsenbein2000}
{Ochsenbein}, F., {Bauer}, P., \& {Marcout}, J. 2000, \aaps, 143, 23,
  \dodoi{10.1051/aas:2000169}

\bibitem[{{Ohlmann} {et~al.}(2014){Ohlmann}, {Kromer}, {Fink}, {Pakmor},
  {Seitenzahl}, {Sim}, \& {R{\"o}pke}}]{ohlmann2014}
{Ohlmann}, S.~T., {Kromer}, M., {Fink}, M., {et~al.} 2014, \aap, 572, A57,
  \dodoi{10.1051/0004-6361/201423924}

\bibitem[{{Papish} \& {Perets}(2016)}]{papish2016}
{Papish}, O., \& {Perets}, H.~B. 2016, \apj, 822, 19,
  \dodoi{10.3847/0004-637X/822/1/19}

\bibitem[{{Queiroz} {et~al.}(2018){Queiroz}, {Anders}, {Santiago}, {Chiappini},
  {Steinmetz}, {Dal Ponte}, {Stassun}, {da Costa}, {Maia}, {Crestani}, {Beers},
  {Fern{\'a}ndez-Trincado}, {Garc{\'\i}a-Hern{\'a}ndez}, {Roman-Lopes}, \&
  {Zamora}}]{queiroz2018}
{Queiroz}, A.~B.~A., {Anders}, F., {Santiago}, B.~X., {et~al.} 2018, \mnras,
  476, 2556, \dodoi{10.1093/mnras/sty330}

\bibitem[{{Queiroz} {et~al.}(2020){Queiroz}, {Anders}, {Chiappini},
  {Khalatyan}, {Santiago}, {Steinmetz}, {Valentini}, {Miglio}, {Bossini},
  {Barbuy}, {Minchev}, {Minniti}, {Garc{\'\i}a Hern{\'a}ndez}, {Schultheis},
  {Beaton}, {Beers}, {Bizyaev}, {Brownstein}, {Cunha},
  {Fern{\'a}ndez-Trincado}, {Frinchaboy}, {Lane}, {Majewski}, {Nataf},
  {Nitschelm}, {Pan}, {Roman-Lopes}, {Sobeck}, {Stringfellow}, \&
  {Zamora}}]{queiroz2020}
{Queiroz}, A.~B.~A., {Anders}, F., {Chiappini}, C., {et~al.} 2020, \aap, 638,
  A76, \dodoi{10.1051/0004-6361/201937364}

\bibitem[{{R Core Team}(2020)}]{r20}
{R Core Team}. 2020, R: A Language and Environment for Statistical Computing, R
  Foundation for Statistical Computing, Vienna, Austria.
\newblock \url{https://www.R-project.org/}

\bibitem[{{Ram{\'\i}rez} {et~al.}(2014){Ram{\'\i}rez}, {Mel{\'e}ndez}, {Bean},
  {Asplund}, {Bedell}, {Monroe}, {Casagrande}, {Schirbel}, {Dreizler}, {Teske},
  {Tucci Maia}, {Alves-Brito}, \& {Baumann}}]{ramirez2014}
{Ram{\'\i}rez}, I., {Mel{\'e}ndez}, J., {Bean}, J., {et~al.} 2014, \aap, 572,
  A48, \dodoi{10.1051/0004-6361/201424244}

\bibitem[{{Reggiani} {et~al.}(2017){Reggiani}, {Mel{\'e}ndez}, {Kobayashi},
  {Karakas}, \& {Placco}}]{reggiani2017}
{Reggiani}, H., {Mel{\'e}ndez}, J., {Kobayashi}, C., {Karakas}, A., \&
  {Placco}, V. 2017, \aap, 608, A46, \dodoi{10.1051/0004-6361/201730750}

\bibitem[{{Roederer} {et~al.}(2014){Roederer}, {Preston}, {Thompson},
  {Shectman}, {Sneden}, {Burley}, \& {Kelson}}]{roederer2014}
{Roederer}, I.~U., {Preston}, G.~W., {Thompson}, I.~B., {et~al.} 2014, \aj,
  147, 136, \dodoi{10.1088/0004-6256/147/6/136}

\bibitem[{{Salvadori} {et~al.}(2010){Salvadori}, {Ferrara}, {Schneider},
  {Scannapieco}, \& {Kawata}}]{salvadori2010}
{Salvadori}, S., {Ferrara}, A., {Schneider}, R., {Scannapieco}, E., \&
  {Kawata}, D. 2010, \mnras, 401, L5, \dodoi{10.1111/j.1745-3933.2009.00772.x}

\bibitem[{{Scannapieco} {et~al.}(2006){Scannapieco}, {Kawata}, {Brook},
  {Schneider}, {Ferrara}, \& {Gibson}}]{scannapieco2006}
{Scannapieco}, E., {Kawata}, D., {Brook}, C.~B., {et~al.} 2006, \apj, 653, 285,
  \dodoi{10.1086/508487}

\bibitem[{{Schlaufman} \& {Casey}(2014)}]{schlaufman2014}
{Schlaufman}, K.~C., \& {Casey}, A.~R. 2014, \apj, 797, 13,
  \dodoi{10.1088/0004-637X/797/1/13}

\bibitem[{{Seitenzahl} {et~al.}(2013{\natexlab{a}}){Seitenzahl}, {Cescutti},
  {R{\"o}pke}, {Ruiter}, \& {Pakmor}}]{seitenzahl2013a}
{Seitenzahl}, I.~R., {Cescutti}, G., {R{\"o}pke}, F.~K., {Ruiter}, A.~J., \&
  {Pakmor}, R. 2013{\natexlab{a}}, \aap, 559, L5,
  \dodoi{10.1051/0004-6361/201322599}

\bibitem[{{Seitenzahl} \& {Townsley}(2017)}]{seitenzahl2017}
{Seitenzahl}, I.~R., \& {Townsley}, D.~M. 2017, {Nucleosynthesis in
  Thermonuclear Supernovae}, ed. A.~W. {Alsabti} \& P.~{Murdin}, 1955,
  \dodoi{10.1007/978-3-319-21846-5_87}

\bibitem[{{Seitenzahl} {et~al.}(2013{\natexlab{b}}){Seitenzahl},
  {Ciaraldi-Schoolmann}, {R{\"o}pke}, {Fink}, {Hillebrand t}, {Kromer},
  {Pakmor}, {Ruiter}, {Sim}, \& {Taubenberger}}]{seitenzahl2013b}
{Seitenzahl}, I.~R., {Ciaraldi-Schoolmann}, F., {R{\"o}pke}, F.~K., {et~al.}
  2013{\natexlab{b}}, \mnras, 429, 1156, \dodoi{10.1093/mnras/sts402}

\bibitem[{{Seitenzahl} {et~al.}(2016){Seitenzahl}, {Kromer}, {Ohlmann},
  {Ciaraldi-Schoolmann}, {Marquardt}, {Fink}, {Hillebrandt}, {Pakmor},
  {R{\"o}pke}, {Ruiter}, {Sim}, \& {Taubenberger}}]{seitenzahl2016}
{Seitenzahl}, I.~R., {Kromer}, M., {Ohlmann}, S.~T., {et~al.} 2016, \aap, 592,
  A57, \dodoi{10.1051/0004-6361/201527251}

\bibitem[{{Sharma} {et~al.}(2018){Sharma}, {Theuns}, \& {Frenk}}]{sharma2018}
{Sharma}, M., {Theuns}, T., \& {Frenk}, C. 2018, \mnras, 479, 1638,
  \dodoi{10.1093/mnras/sty1319}

\bibitem[{{Shectman} \& {Johns}(2003)}]{shectman2003}
{Shectman}, S.~A., \& {Johns}, M. 2003, in Society of Photo-Optical
  Instrumentation Engineers (SPIE) Conference Series, Vol. 4837, \procspie, ed.
  J.~M. {Oschmann} \& L.~M. {Stepp}, 910--918, \dodoi{10.1117/12.457909}

\bibitem[{{Singh} {et~al.}(2020){Singh}, {Hansen}, {Byrgesen}, {Reichert}, \&
  {Reggiani}}]{singh2020}
{Singh}, D., {Hansen}, C.~J., {Byrgesen}, J.~S., {Reichert}, M., \& {Reggiani},
  H.~M. 2020, \aap, 634, A72, \dodoi{10.1051/0004-6361/201936305}

\bibitem[{{Skrutskie} {et~al.}(2006){Skrutskie}, {Cutri}, {Stiening},
  {Weinberg}, {Schneider}, {Carpenter}, {Beichman}, {Capps}, {Chester},
  {Elias}, {Huchra}, {Liebert}, {Lonsdale}, {Monet}, {Price}, {Seitzer},
  {Jarrett}, {Kirkpatrick}, {Gizis}, {Howard}, {Evans}, {Fowler}, {Fullmer},
  {Hurt}, {Light}, {Kopan}, {Marsh}, {McCallon}, {Tam}, {Van Dyk}, \&
  {Wheelock}}]{skrutskie2006}
{Skrutskie}, M.~F., {Cutri}, R.~M., {Stiening}, R., {et~al.} 2006, \aj, 131,
  1163, \dodoi{10.1086/498708}

\bibitem[{{Sneden} {et~al.}(2009){Sneden}, {Lawler}, {Cowan}, {Ivans}, \& {Den
  Hartog}}]{sneden2009}
{Sneden}, C., {Lawler}, J.~E., {Cowan}, J.~J., {Ivans}, I.~I., \& {Den Hartog},
  E.~A. 2009, \apjs, 182, 80, \dodoi{10.1088/0067-0049/182/1/80}

\bibitem[{{Sneden} {et~al.}(2016){Sneden}, {Lawler}, {den Hartog}, \&
  {Wood}}]{sneden2016}
{Sneden}, C., {Lawler}, J.~E., {den Hartog}, E.~A., \& {Wood}, M.~E. 2016, IAU
  Focus Meeting, 29A, 287, \dodoi{10.1017/S1743921316003069}

\bibitem[{{Sneden}(1973)}]{sneden1973}
{Sneden}, C.~A. 1973, PhD thesis, THE UNIVERSITY OF TEXAS AT AUSTIN.

\bibitem[{{Starkenburg} {et~al.}(2017){Starkenburg}, {Oman}, {Navarro},
  {Crain}, {Fattahi}, {Frenk}, {Sawala}, \& {Schaye}}]{starkenburg2017b}
{Starkenburg}, E., {Oman}, K.~A., {Navarro}, J.~F., {et~al.} 2017, \mnras, 465,
  2212, \dodoi{10.1093/mnras/stw2873}

\bibitem[{{Tody}(1986)}]{iraf1986}
{Tody}, D. 1986, Society of Photo-Optical Instrumentation Engineers (SPIE)
  Conference Series, Vol. 627, {The IRAF Data Reduction and Analysis System},
  ed. D.~L. {Crawford}, 733, \dodoi{10.1117/12.968154}

\bibitem[{{Tody}(1993)}]{iraf1993}
---. 1993, Astronomical Society of the Pacific Conference Series, Vol.~52,
  {IRAF in the Nineties}, ed. R.~J. {Hanisch}, R.~J.~V. {Brissenden}, \&
  J.~{Barnes}, 173

\bibitem[{{Tumlinson}(2010)}]{tumlinson2010}
{Tumlinson}, J. 2010, \apj, 708, 1398, \dodoi{10.1088/0004-637X/708/2/1398}

\bibitem[{{van der Walt} {et~al.}(2011){van der Walt}, {Colbert}, \&
  {Varoquaux}}]{vanderwalt2011}
{van der Walt}, S., {Colbert}, S.~C., \& {Varoquaux}, G. 2011, Computing in
  Science and Engineering, 13, 22, \dodoi{10.1109/MCSE.2011.37}

\bibitem[{{Virtanen} {et~al.}(2020){Virtanen}, {Gommers}, {Oliphant},
  {Haberland}, {Reddy}, {Cournapeau}, {Burovski}, {Peterson}, {Weckesser},
  {Bright}, {van der Walt}, {Brett}, {Wilson}, {Jarrod Millman}, {Mayorov},
  {Nelson}, {Jones}, {Kern}, {Larson}, {Carey}, {Polat}, {Feng}, {Moore}, {Vand
  erPlas}, {Laxalde}, {Perktold}, {Cimrman}, {Henriksen}, {Quintero}, {Harris},
  {Archibald}, {Ribeiro}, {Pedregosa}, {van Mulbregt}, \&
  {Contributors}}]{scipy2020}
{Virtanen}, P., {Gommers}, R., {Oliphant}, T.~E., {et~al.} 2020, Nature
  Methods, 17, 261, \dodoi{https://doi.org/10.1038/s41592-019-0686-2}

\bibitem[{{Wang} {et~al.}(2009{\natexlab{a}}){Wang}, {Chen}, {Meng}, \&
  {Han}}]{wang2009a}
{Wang}, B., {Chen}, X., {Meng}, X., \& {Han}, Z. 2009{\natexlab{a}}, \apj, 701,
  1540, \dodoi{10.1088/0004-637X/701/2/1540}

\bibitem[{{Wang} {et~al.}(2009{\natexlab{b}}){Wang}, {Meng}, {Chen}, \&
  {Han}}]{wang2009b}
{Wang}, B., {Meng}, X., {Chen}, X., \& {Han}, Z. 2009{\natexlab{b}}, \mnras,
  395, 847, \dodoi{10.1111/j.1365-2966.2009.14545.x}

\bibitem[{{Wenger} {et~al.}(2000){Wenger}, {Ochsenbein}, {Egret}, {Dubois},
  {Bonnarel}, {Borde}, {Genova}, {Jasniewicz}, {Lalo{\"e}}, {Lesteven}, \&
  {Monier}}]{wenger2000}
{Wenger}, M., {Ochsenbein}, F., {Egret}, D., {et~al.} 2000, \aaps, 143, 9,
  \dodoi{10.1051/aas:2000332}

\bibitem[{{W}es {M}c{K}inney(2010)}]{scipy2010}
{W}es {M}c{K}inney. 2010, in {P}roceedings of the 9th {P}ython in {S}cience
  {C}onference, ed. {S}t\'efan van~der {W}alt \& {J}arrod {M}illman, 56 -- 61,
  \dodoi{10.25080/Majora-92bf1922-00a}

\bibitem[{{Wolf} {et~al.}(2018){Wolf}, {Onken}, {Luvaul}, {Schmidt}, {Bessell},
  {Chang}, {Da Costa}, {Mackey}, {Martin-Jones}, {Murphy}, {Preston}, {Scalzo},
  {Shao}, {Smillie}, {Tisserand}, {White}, \& {Yuan}}]{wolf2018}
{Wolf}, C., {Onken}, C.~A., {Luvaul}, L.~C., {et~al.} 2018, \pasa, 35, e010,
  \dodoi{10.1017/pasa.2018.5}

\bibitem[{{Woosley} \& {Weaver}(1994)}]{woosley1994}
{Woosley}, S.~E., \& {Weaver}, T.~A. 1994, \apj, 423, 371,
  \dodoi{10.1086/173813}

\bibitem[{{Woosley} \& {Weaver}(1995)}]{woosley1995}
---. 1995, \apjs, 101, 181, \dodoi{10.1086/192237}

\bibitem[{{Wright}(2006)}]{wright2006}
{Wright}, E.~L. 2006, \pasp, 118, 1711, \dodoi{10.1086/510102}

\bibitem[{{Wright} {et~al.}(2010){Wright}, {Eisenhardt}, {Mainzer}, {Ressler},
  {Cutri}, {Jarrett}, {Kirkpatrick}, {Padgett}, {McMillan}, {Skrutskie},
  {Stanford}, {Cohen}, {Walker}, {Mather}, {Leisawitz}, {Gautier}, {McLean},
  {Benford}, {Lonsdale}, {Blain}, {Mendez}, {Irace}, {Duval}, {Liu}, {Royer},
  {Heinrichsen}, {Howard}, {Shannon}, {Kendall}, {Walsh}, {Larsen}, {Cardon},
  {Schick}, {Schwalm}, {Abid}, {Fabinsky}, {Naes}, \& {Tsai}}]{wright2010}
{Wright}, E.~L., {Eisenhardt}, P. R.~M., {Mainzer}, A.~K., {et~al.} 2010, \aj,
  140, 1868, \dodoi{10.1088/0004-6256/140/6/1868}

\bibitem[{{Yamaguchi} {et~al.}(2015){Yamaguchi}, {Badenes}, {Foster}, {Bravo},
  {Williams}, {Maeda}, {Nobukawa}, {Eriksen}, {Brickhouse}, {Petre}, \&
  {Koyama}}]{yamaguchi2015}
{Yamaguchi}, H., {Badenes}, C., {Foster}, A.~R., {et~al.} 2015, \apjl, 801,
  L31, \dodoi{10.1088/2041-8205/801/2/L31}

\end{thebibliography}
\bibliographystyle{aasjournal}
\listofchanges

\end{document}